\documentclass[twoside,twocolumn,9pt]{article}
\usepackage{extsizes}
\usepackage[super,sort&compress,comma]{natbib} 
\usepackage[version=3]{mhchem}
\usepackage[left=1.5cm, right=1.5cm, top=1.785cm, bottom=2.0cm]{geometry}
\usepackage{balance}
\usepackage{mathptmx}
\usepackage{sectsty}
\usepackage{graphicx} 
\usepackage{lastpage}
\usepackage[format=plain,justification=justified,singlelinecheck=false,font={stretch=1.125,small,sf},labelfont=bf,labelsep=space]{caption}
\usepackage{float}
\usepackage{fancyhdr}
\usepackage{fnpos}
\usepackage[english]{babel}
\addto{\captionsenglish}{%
  
}
\usepackage{array}
\usepackage{droidsans}
\usepackage{charter}
\usepackage[T1]{fontenc}
\usepackage[usenames,dvipsnames]{xcolor}
\usepackage{setspace}
\usepackage[compact]{titlesec}
\usepackage[colorlinks=false]{hyperref}
\usepackage{siunitx}
\usepackage{verbatim}
\usepackage{footnote}

\newcommand*{\wn}{cm$^{-1}$}
\newcommand*{\Tm}{T$_{2}$}
\newcommand*{\Hm}{H$_{2}$}
\newcommand*{\Dm}{D$_{2}$}

\newcolumntype{d}{D{.}{.}{-1}}
\definecolor{cream}{RGB}{222,217,201}

\begin{document}

\pagestyle{fancy}
\thispagestyle{plain}
\fancypagestyle{plain}{
\renewcommand{\headrulewidth}{0pt}
}

\makeFNbottom
\makeatletter
\renewcommand\LARGE{\@setfontsize\LARGE{15pt}{17}}
\renewcommand\Large{\@setfontsize\Large{12pt}{14}}
\renewcommand\large{\@setfontsize\large{10pt}{12}}
\renewcommand\footnotesize{\@setfontsize\footnotesize{7pt}{10}}
\makeatother

\renewcommand{\thefootnote}{\fnsymbol{footnote}}
\renewcommand\footnoterule{\vspace*{1pt}%
\color{cream}\hrule width 3.5in height 0.4pt \color{black}\vspace*{5pt}} 
\setcounter{secnumdepth}{5}

\makeatletter 
\renewcommand\@biblabel[1]{#1}            
\renewcommand\@makefntext[1]%
{\noindent\makebox[0pt][r]{\@thefnmark\,}#1}
\makeatother 
\renewcommand{\figurename}{\small{Fig.}~}
\sectionfont{\sffamily\Large}
\subsectionfont{\normalsize}
\subsubsectionfont{\bf}
\setstretch{1.125} 
\setlength{\skip\footins}{0.8cm}
\setlength{\footnotesep}{0.25cm}
\setlength{\jot}{10pt}
\titlespacing*{\section}{0pt}{4pt}{4pt}
\titlespacing*{\subsection}{0pt}{15pt}{1pt}

\fancyfoot{}
\fancyfoot[LO,RE]{\vspace{-7.1pt}\includegraphics[height=9pt]{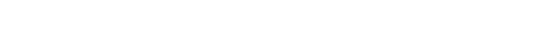}}
\fancyfoot[CO]{\vspace{-7.1pt}\hspace{11.9cm}\includegraphics{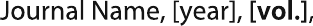}}
\fancyfoot[CE]{\vspace{-7.2pt}\hspace{-13.2cm}\includegraphics{head_foot/RF}}
\fancyfoot[RO]{\footnotesize{\sffamily{1--\pageref{LastPage} ~\textbar  \hspace{2pt}\thepage}}}
\fancyfoot[LE]{\footnotesize{\sffamily{\thepage~\textbar\hspace{4.65cm} 1--\pageref{LastPage}}}}
\fancyhead{}
\renewcommand{\headrulewidth}{0pt} 
\renewcommand{\footrulewidth}{0pt}
\setlength{\arrayrulewidth}{1pt}
\setlength{\columnsep}{6.5mm}
\setlength\bibsep{1pt}

\makeatletter 
\newlength{\figrulesep} 
\setlength{\figrulesep}{0.5\textfloatsep} 

\newcommand{\topfigrule}{\vspace*{-1pt}%
\noindent{\color{cream}\rule[-\figrulesep]{\columnwidth}{1.5pt}} }

\newcommand{\botfigrule}{\vspace*{-2pt}%
\noindent{\color{cream}\rule[\figrulesep]{\columnwidth}{1.5pt}} }

\newcommand{\dblfigrule}{\vspace*{-1pt}%
\noindent{\color{cream}\rule[-\figrulesep]{\textwidth}{1.5pt}} }

\makeatother

\twocolumn[
  \begin{@twocolumnfalse}
{\includegraphics[height=30pt]{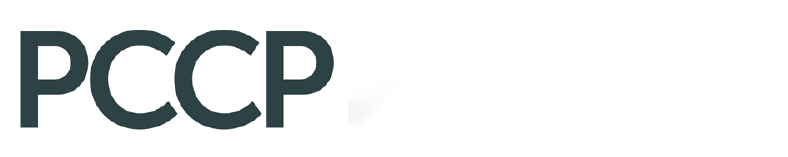}\hfill\raisebox{0pt}[0pt][0pt]{\includegraphics[height=55pt]{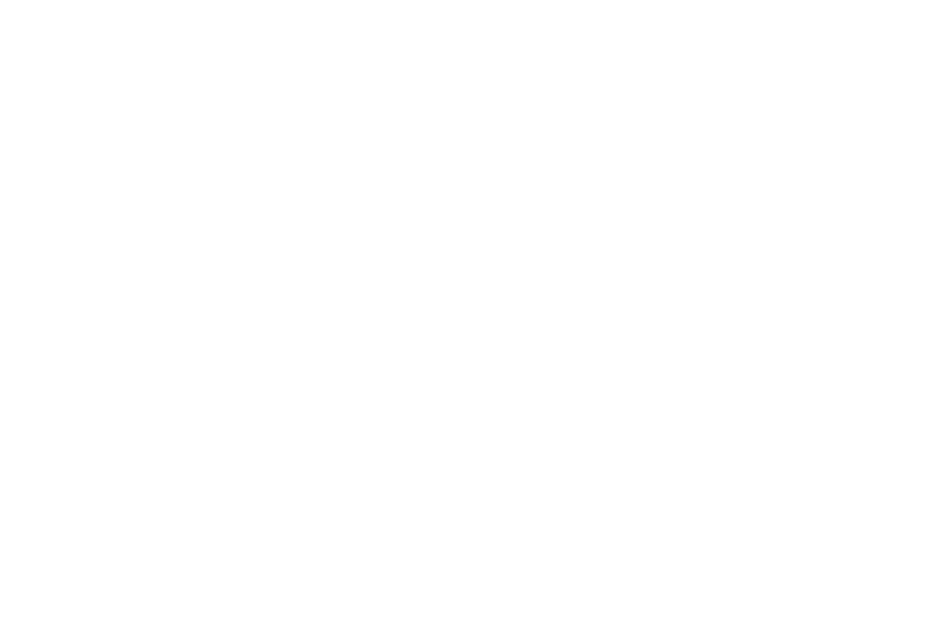}}\\[1ex]
\includegraphics[width=18.5cm]{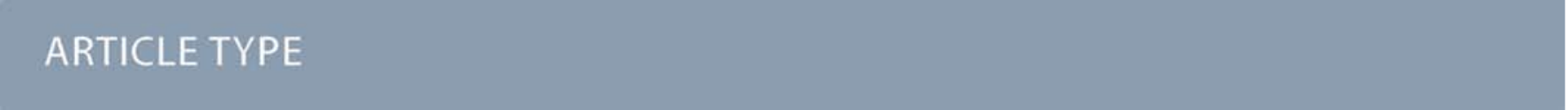}}\par
\vspace{1em}
\sffamily
\begin{tabular}{m{4.5cm} p{13.5cm} }

\includegraphics{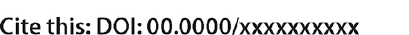} & \noindent\LARGE{\textbf{Precision measurement of the fundamental vibrational frequencies of tritium-bearing hydrogen molecules: T$_2$, DT, HT}} \\
\vspace{0.3cm} & \vspace{0.3cm} \\

 & \noindent\large{
 K.-F. Lai,\textit{$^{a}$} 
 V. Hermann,\textit{$^{b}$} 
 T. M. Trivikram,\textit{$^{a}$}$^{\dag}$
 M. Diouf,\textit{$^{a}$}
 M. Schl\"{o}sser,\textit{$^{b}$} 
 W. Ubachs,\textit{$^{a}$} and 
 E. J. Salumbides\textit{$^{a}$}$^{\ddag}$
 } \\

\includegraphics{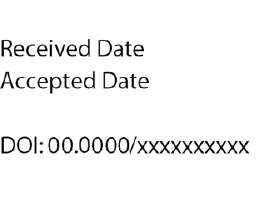} & \noindent\normalsize{High-resolution coherent Raman spectroscopic measurements of all three tritium-containing molecular hydrogen isotopologues \Tm, DT and HT were performed to determine the ground electronic state fundamental Q-branch ($v=0 \rightarrow 1, \Delta J = 0$) transition frequencies at accuracies of $0.0005$ \wn. An over hundred-fold improvement in accuracy over previous experiments allows the comparison with the latest ab initio calculations in the framework of Non-Adiabatic Perturbation Theory including nonrelativisitic, relativisitic and QED contributions. Excellent agreement is found between experiment and theory, thus providing a verification of the validity of the  NAPT-framework for these tritiated species. While the transition frequencies were corrected for ac-Stark shifts, the contributions of non-resonant background as well as quantum interference effects between resonant features in the nonlinear spectroscopy were quantitatively investigated, also leading to corrections to the transition frequencies. Methods of saturated CARS with the observation of Lamb dips, as well as the use of continuous-wave radiation for the Stokes frequency were explored, that might pave the way for future higher-accuracy CARS measurements.
}

\end{tabular}

 \end{@twocolumnfalse} \vspace{0.6cm}

]

\renewcommand*\rmdefault{bch}\normalfont\upshape
\rmfamily
\section*{}
\vspace{-1cm}


\footnotetext{\textit{$^{a}$Department of Physics and Astronomy, LaserLaB, Vrije Universiteit Amsterdam, De Boelelaan 1081, 1081 HV Amsterdam, The Netherlands}}
\footnotetext{\textit{$^{b}$Tritium Laboratory Karlsruhe, Institute of Nuclear Physics, Karlsruhe Institute of Technology, Hermann-von-Helmholtz-Platz 1, 76344 Eggenstein-Leopoldshafen, Germany}}

\footnotetext{\textit{\dag~Present address: Max Born Institute, Max-Born-Straße 2A, 12489 Berlin, Germany}}
\footnotetext{\textit{\ddag~Email: e.j.salumbides@vu.nl}}


\section{Introduction}

Molecular hydrogen represents a benchmark molecule for testing quantum chemical theories. The comparison between precision spectroscopic measurements on the level structure of hydrogen and the results from calculations has made \Hm\ and its isotopologues a test bed for searches of fifth forces~\cite{Salumbides2013}, higher dimensions~\cite{Salumbides2015b}, and physics beyond the Standard Model of physics~\cite{Ubachs2016}.
On the experimental side the measurement of the dissociation energy of \Hm\ has witnessed great improvements in the past decade~\cite{Liu2009,Cheng2018,Holsch2019,Beyer2019} now reaching a relative accuracy of $3\times10^{-10}$. Each isotopologue supports over 300 bound and long-lived rovibrational levels,  that provide an extended playing field for performing fundamental tests,  such as the ground tone vibrational splitting \cite{Dickenson2013}, the quadrupole overtone transitions of \Hm~\cite{Campargue2012,Kassi2014,Cheng2012} and in the \Dm\ species~\cite{Maddaloni2010,Kassi2012}, as well as in the mixed isotopologue HD~\cite{Kassi2011,Tao2018,Cozijn2018,Fasci2018}.

On the theoretical side level energy calculations of the molecular hydrogen four-body system have undergone equally great improvements, producing highly accurate level energies of H$_2$ and D$_2$~\cite{Komasa2011} as well as for HD~\cite{Pachucki2010b}. These methods were based on calculations of Born-Oppenheimer energies~\cite{Pachucki2010}, and separate approaches for adiabatic~\cite{Pachucki2014} and non-adiabatic~\cite{Pachucki2015} corrections, as well as computations of relativistic~\cite{Puchalski2017} and quantum electrodynamic effects~\cite{Puchalski2016}.
This development has led to the comprehensive approach of nonadiabatic perturbation theory (NAPT)~\cite{Czachorowski2018,Komasa2019}, which now enables the rapid computation of all bound rovibrational states in the ground electronic manifold of the hydrogen isotopologues.
In addition, and alongside, an even more refined and more accurate method involving direct 4-body calculations is being pursued, but this is currently limited to the computation of the binding energy of the lowest rovibrational level~\cite{Pachucki2018,Puchalski2019}.

Most of the experimental precision spectroscopic studies were performed in the typical environment of molecular beams that can be easily applied to the stable, non-radioactive, hydrogen isotopologues \Hm, HD and \Dm. Tritium-containing isotopologues, HT, DT and \Tm\ have been much less frequently studied, because of limited access and handling difficulties due to safety requirements holding for radioactive species. The first spectroscopic studies on tritium bearing hydrogen molecules were performed by Dieke and Tomkins at Argonne National Laboratories recording emission spectra~\cite{Dieke1949,Dieke1958}. Later, other methods were applied, such as spontaneous Raman spectroscopy~\cite{Edwards1978,Edwards1979,Veirs1987} and intracavity laser absorption, combined with an optophone~\cite{Chuang1987}.
A rationale for exploring tritium spectroscopy is that the heavier, tritiated molecular hydrogen species exhibit smaller non-adiabatic contributions to the level energies. High precision studies on these species may help to disentangle the effect of mass-dependent terms in the calculation of hydrogen level energies. Also, inclusion of tritium provides two additional heteronuclear species, HT and DT.  These species provide information on the $g/u$ mixing effects present in heteronuclear species. Such studies might help resolve the discrepancy observed in HD, between the experimental value for the dissociation energy~\cite{Sprecher2010} and the theoretical value~\cite{Puchalski2019}. 

The precise level structure of the tritium bearing hydrogen molecules is of relevance for the Karlsruhe Tritium Neutrino experiment (KATRIN) focusing on the $\beta$ decay from \Tm~\cite{Aker2019}. The $\beta$-electron energy spectrum depends on the final state distribution of the $^{3}$HeT$^{+}$ daughter molecules~\cite{Saenz2000}, requiring accurate level energies of $^{3}$HeT$^{+}$, but also on the isotopologue distribution, parent quantum state distribution, and the level structure of the parent states~\cite{Doss2006,Bodine2015,Kleesiek2019}. The Raman spectroscopic techniques for molecular hydrogen isotopologues, including tritium-bearing species, were further developed to determine composition and concentration of the isotope mixtures for the KATRIN experiment~\cite{James2013,Schlosser2013,Schloesser2015}.

Here we present a study of transition frequencies in the fundamental vibrational band of all tritiated species, \Tm, DT and HT, via Coherent Anti-Stokes Raman Scattering (CARS). After preliminary reports on the CARS detection~\cite{Schlosser2017}, on the spectroscopy of the \Tm\ isotopologue~\cite{Trivikram2018} and DT~\cite{Lai2019}, we now present a comprehensive report, including results on HT and further detailing the experimental methods. It is noted that remeasurements were performed for the \Tm\ species yielding improved statistics, following up on a previous set of data~\cite{Trivikram2018}. A comparison is made with the findings of theoretical calculations in the NAPT-framework~\cite{Komasa2019}.

\section{Experiment}

\begin{figure}
\begin{center}
\includegraphics[width=1\linewidth]{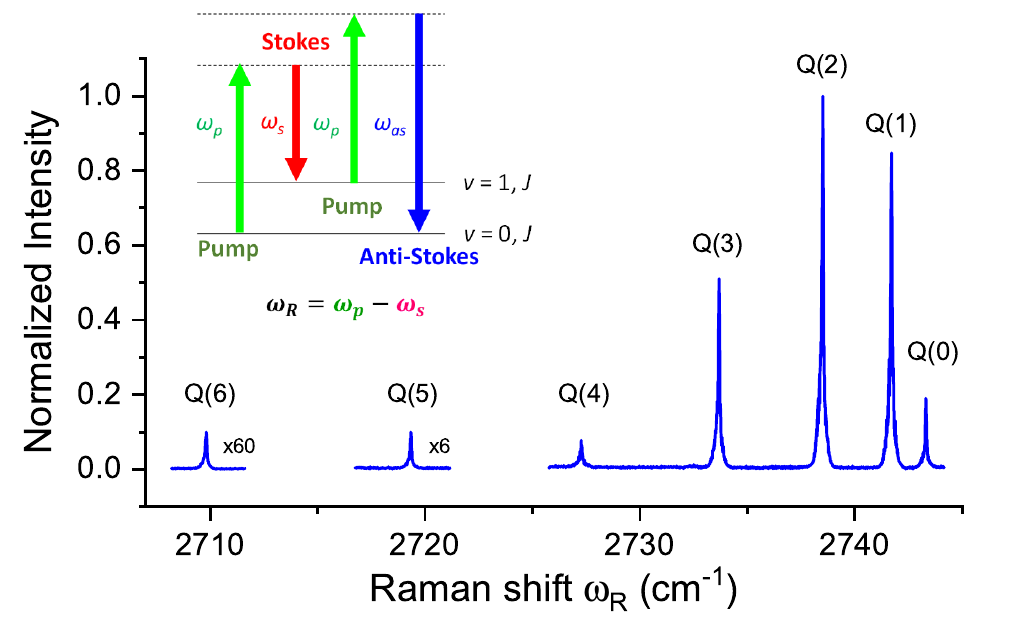}
\caption{\label{broadband}
Overview recording of the $X\, \Sigma^+_g \, (v=0\rightarrow 1)$ Q(0 - 6) lines in DT using a combination of an injection-seeded Nd:YAG laser (pump) and a grating-based pulsed-dye laser (Stokes). The inset shows an energy diagram for the CARS four-wave mixing scheme, with the generated Anti-Stokes radiation detected as signal.
}
\end{center}
\end{figure}

For designing an experiment on tritium-bearing hydrogen isotopologues aspects of obtainable experimental precision have to be matched to safety regulations with regard to radioactivity. The requirement of a legal upper limit of 1 GBq of tritium stored in the laboratory, corresponding in case of \Tm\ to a gas amount of $11.5$ mbar$\cdot$cm$^3$ at room temperature, and the prohibition of pumping gas into an exhaust, sets constraints on the experimental method.  Whereas in many previous precision studies on molecular hydrogen multi-step laser excitation was combined with ion detection~\cite{Liu2009,Altmann2018,Cheng2018,Holsch2019} this is not possible in case of tritium, in view of the $\sim 10^9$ fast electrons and ions being produced per second from the radioactive sample. In view of containment requirements molecular beam studies are ruled out as well. 

A closed-cell environment with a fill of static gas is chosen as a safe solution. Spectroscopic detection via coherent Raman processes is more sensitive than direct linear infra-red absorption, while in both cases Doppler broadening scales linearly with the vibrational energy, so imposing the same level of frequency uncertainty. These arguments led us to the choice of performing CARS experiments for the tritium-bearing hydrogen molecules.
The CARS nonlinear optical process generates a coherent beam of light propagating into the far field where the Anti-Stokes signal can be collected and detected.
In this section the handling of the tritium gas cell, the optical layout of the CARS setup, and the frequency calibration are detailed, while systematic effects are discussed.

\subsection{Tritium gas samples}

The sample cell for the present CARS measurements is specially built from well-proven materials for tritium gas containment, and has been described previously~\cite{Schlosser2017}. The cell with an inner volume of $4\,\mathrm{cm}^3$ has input and output windows placed at a distance of $8\,\mathrm{cm}$, so that focusing at $f=20\,\mathrm{cm}$ is possible without burning window surfaces and avoiding a too high intensity in the focus, in view of issues related to the ac-Stark effect (see below). 

First, a high purity \Tm\ gas sample was prepared in the Tritium Laboratory Karlsruhe (TLK). The cell filled was filled to $2.5\,\mathrm{mbar}$ to be below the legal activity limit including a safety margin.

The heteronuclear molecules HT or DT can be generated by catalytic self-exchange reactions driven by ions\cite{Sazonov2011} produced in the constant $\beta$-radiation intensity from tritium decay.
\[
\text{X}_2 \, + \, \text{T}_2\,  \rightleftharpoons \, 2 \text{XT}, \]
with X being H or D.
The estimated HT and DT partial pressures are derived from equilibrium constants $K_{eq}$:
\[
K_{eq} = \frac{[\text{XT}]^2}{[\text{X}_2] [\text{T}_2]},
\]
Values for $K_{eq}$ are determined~\cite{Jones1967} at 2.58 for HT and 3.82 for DT at $300\,\mathrm{K}$.

For the preparation of DT, first \Dm\ and \Tm\ were injected at a 4:1 ratio into a large mixing vessel. After a short time DT is formed via radio-chemical equilibration. A sample was extracted and filled into the CARS cell up to a total pressure of 12.7 mbar. The partial pressure of DT is estimated to be about 3.8 mbar.
At the time of the preparation of the HT cell, this accurate mixing procedure was not available due to constraints with other running experiments. Thus, HT preparation needed to be performed inside the CARS-cell. The 2.5 mbar \Tm\ sample was topped up with \Hm\ up to a total pressure of 11.5 mbar. This would lead to a theoretical partial pressure of 3.6 mbar. It should be noted that this procedure is less accurately controllable as tritium gas might have been partially back-diffused from the cell reducing the actual HT yield. 

In view of the tritium radioactive half-life of 12.3 years these (partial) pressures effectively remained constant during the measurements performed in time windows of months. The safety-controlled and tritium-gas filled cell was transported to LaserLaB Amsterdam, where the CARS measurements were carried out.

\begin{figure}
\begin{center}
\includegraphics[width=1\linewidth]{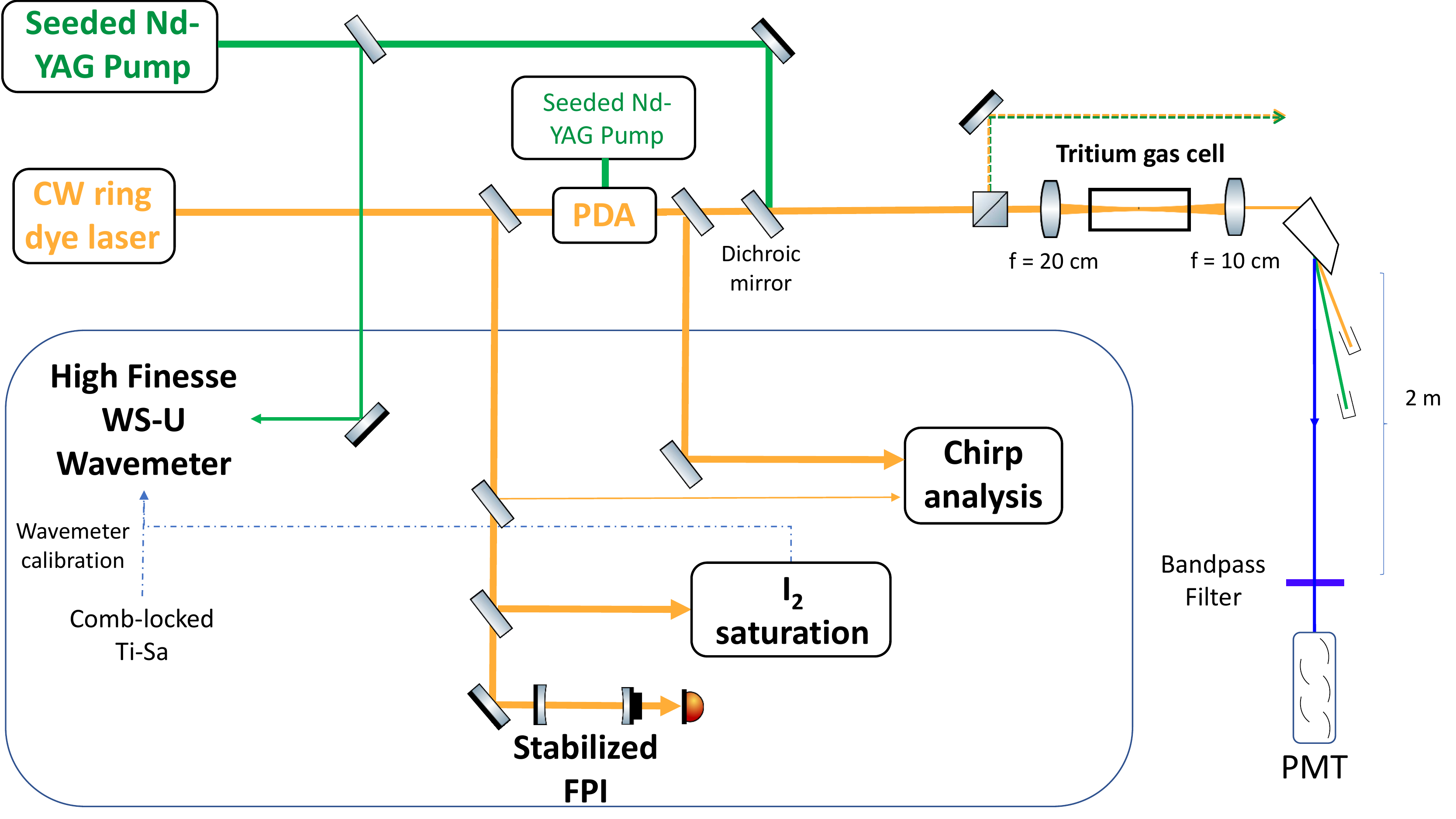}
\caption{\label{setup}
Optical layout for the CARS spectroscopy setup. The frame encloses the setups for frequency calibration of the pump and Stokes beams. See text for more details.}
\end{center}
\end{figure}

\subsection{Precision CARS setup}

In the present study, the fundamental band ($\nu = 1 \gets 0$) Q-branch transitions of all three tritiated species are measured by the CARS non-linear optical technique. The specific resonant four-wave mixing scheme of the CARS process is shown in Fig.~\ref{broadband} alongside with an overview spectrum of the Q-lines in DT. An anti-Stokes field is generated, at frequency $\omega_{AS} = 2\omega_{P} - \omega_{S}$, using input fields of a pump frequency $\omega_{P}$ and a Stokes frequency $\omega_{S}$. The integrated anti-Stokes signal intensity $I_{AS}$ is produced as a coherent light beam in the forward direction under phase-matching conditions and is proportional to~\cite{Tolles1977}:
\begin{equation} \label{eq:Ias}
I_{AS} \propto \int \int \vert{\chi^{(3)}}\vert^2 I_{P}^{2}(z,t) I_{S}(z,t)dzdt,
\end{equation}
where $I_P(z,t)$ and $I_S(z,t)$ are the spatial and temporal evolution of pump and Stokes intensity and $z$ is the optical path. 
The third order nonlinear susceptibility $\chi^{(3)}$ depends on the molecular number density and the differential Raman cross-section~\cite{Tolles1977}.

The optical layout of the CARS setup is shown in Fig.~\ref{setup}. The pump beam is obtained from a frequency-doubled injection-seeded Nd-YAG laser (Spectra-Physics GCR-330) delivering pulses at 532 nm, of 8 ns duration, which would correspond to a bandwidth of about 55 MHz in case of Fourier transform (FT) limited laser pulses.  During the measurements, only low pulse energies were used, ranging from 30 to 500 $\mu$J in the interaction range. 

For the precision CARS measurements the Stokes beam is obtained from an injection-seeded traveling-wave pulsed dye amplifier (PDA) operating on DCM dye in methanol~\cite{Eikema1997}. The continuous-wave (cw) seed light for the PDA is obtained through an optical fiber from a ring-dye laser, covering the range from 590 nm to 650 nm. With 150 mW seed power and 50 mJ pumping from a second frequency-doubled injection-seeded Nd-YAG laser (Spectra-Physics Pro-250), the output energy reaches about 5 mJ at 633 nm (peak performance wavelength) at 6 ns pulse duration, which would correspond to a bandwidth of about a 74 MHz in case of FT-limited laser pulses; the actual width will be slightly larger in view of the chirp phenomena detected (see below). The pulse energy of the Stokes beam in the CARS experiment was kept unchanged for most part of the measurement campaigns at $< 20$ $\mu$J. 

The pump and Stokes beams are combined on a longpass dichroic mirror and passed through a polarization beam splitter cube to ensure parallel polarizations. The two pulses are focused co-linearly into a gas cell with a $f = 20$ cm lens. The flat-top pump pulse beam waist is about 44 $\mu$m, as measured by a CCD camera, for a 3 mm Rayleigh length. The Stokes pulse is estimated to be 1.7 times larger in diameter than the pump beam. Part of the merged beams is sampled out over a long distance for alignment purposes, with the angle mismatch estimated to be better than 2 mrad. 
The generated CARS signal (at $\lambda \sim 450 - 470$ nm) travels along the same propagation axis of the pump and Stokes beams. The CARS-signal beam is collimated with a $f = 10$ cm lens, dispersed from the incident beams by a Pellin-Broca prism, and propagated over a 2 m separation path. The signal beam is then passed through an aperture and bandpass filter, and detected by photomultiplier tube (Philips XP-1911) mounted inside a dark box.

The measurements are performed with few-mbar pressure gas samples, and the spectra obtained are expected to be Doppler-limited. The expression for Doppler broadening (FWHM) of forward spontaneous Raman scattering is~\cite{Lucht1988}:
\begin{equation} \label{eq:dopplerwidth}
\Delta \nu = \frac{2\nu_{0}}{c}\left(\frac{2k_BT\ln{2}}{m}\right)^{1/2}
\end{equation}
where $\nu_{0}$ is the fundamental vibrational frequency splitting, $m$ is the molecular mass, $k_B$ the Boltzmann constant, $c$ the speed of light, and $T$ the temperature. This results in a Doppler width at room temperature of 370 MHz for the heaviest species \Tm, and about 450 MHz and 630 MHz for DT and HT, respectively.
CARS spectral profile simulations by Lucht and Farrow~\cite{Lucht1988} showed that the spectral width of the CARS resonances is about 1.2-times that of the spontaneous Raman scattering in the forward direction.
These values for the width are to be further increased by contributions of the laser bandwidth of both lasers, some additional broadening caused by the timing jitter in the temporal overlap of the two pulses (within 3.5 ns), and by ac-Stark broadening.

\subsection{Frequency calibration}

The frequency calibration of the Stokes beam was performed in two stages. As shown in Fig.~\ref{setup}, part of the cw seed-radiation was split off to perform I$_2$-saturation spectroscopy for absolute calibration against well-calibrated I$_2$-standards~\cite{Xu2000}, and another part for measuring transmission markers from an etalon (FSR $= 150.67(2)$ MHz) stabilized by a HeNe laser (Thorlabs HRS-015). The resulting uncertainty in the absolute frequency calibration of the cw seed laser amounts to 2 MHz.

\begin{figure}
\begin{center}
\includegraphics[width=1\linewidth]{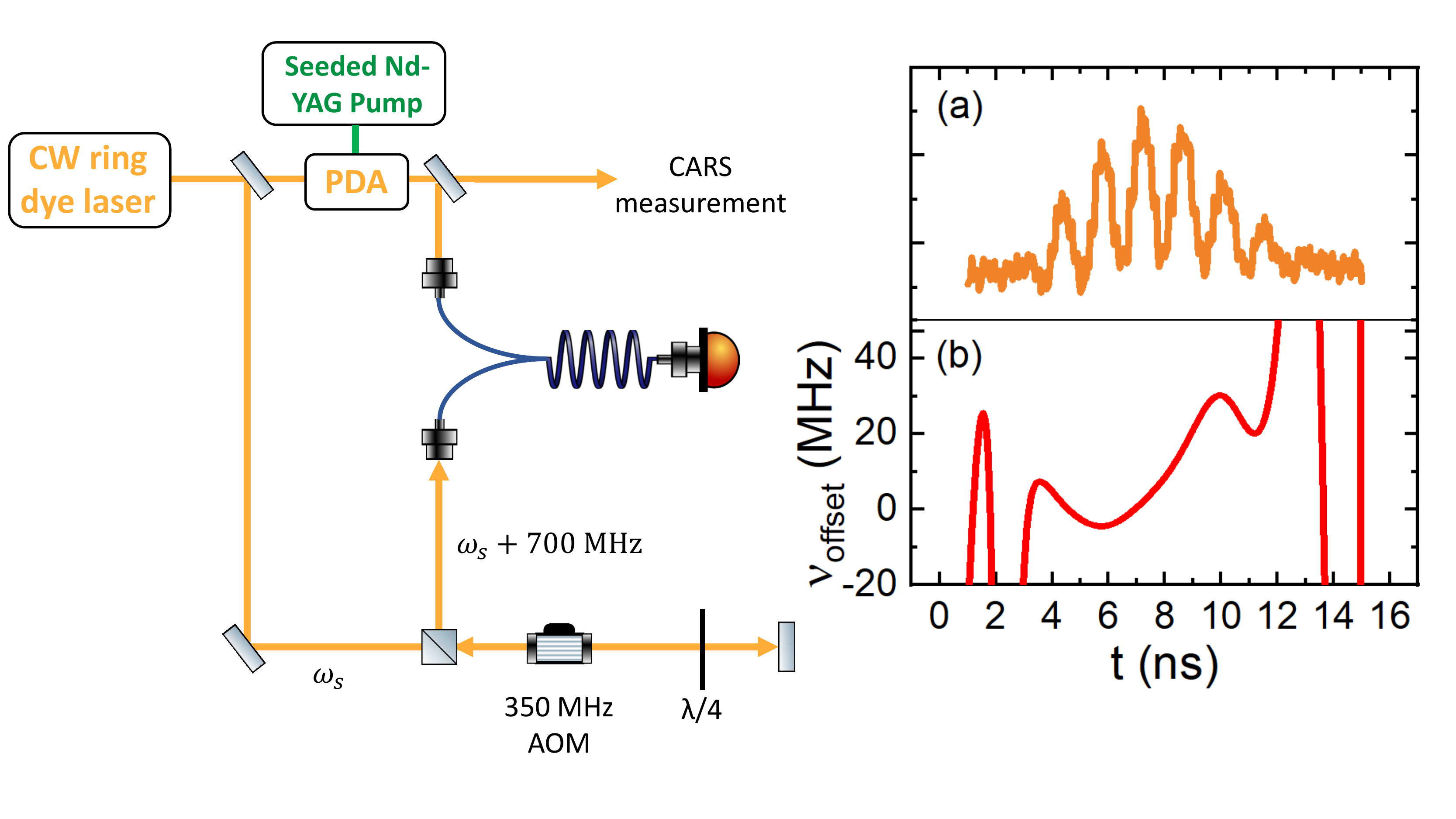}
\caption{\label{chirp}
Setup for the frequency chirp measurement of the output of the PDA laser system. The cw-seed laser is shifted by double-passing an AOM at 350 MHz modulation frequency. The shifted cw and pulsed laser outputs are coupled through a single mode 2 x 2 fused fibre coupler for imposing spatial overlap. The beat note in (a) is detected on a fast photo-detector (Thorlabs DET025AFC) and stored in a Tektronix TDS7404 oscilloscope. b) The instantaneous frequency offset, $\nu_{offset} = \nu_{pulse} - \nu_{cw}$, is obtained from the phase evolution of the 700 MHz carrier signal ~\cite{Hannemann2007a}.
}
\end{center}
\end{figure}

In a second step the chirp-induced pulse-cw frequency offset was measured by techniques previously employed for frequency calibration of a PDA-system~\cite{Fee1992,Gangopadhyay1994,Eikema1997}. The frequency of the cw-seed was shifted by +700 MHz by an acousto-optic modulator (AOM) in a double-pass configuration and the beatnote between the shifted cw-seeding and Stokes pulses was recorded using a fast photodetector and a 4-GHz bandwidth oscilloscope.
The setup for producing a beat-note signal is displayed in Fig.~\ref{chirp}a.
Following Hannemann et al.~\cite{Hannemann2007a} for the frequency chirp analysis, the time-dependent phase in the duration of the pulse is extracted from the beat-note signal, to yield a time-dependent frequency offset curve as plotted in Fig.~\ref{chirp}b.
The reported average frequency offset values are obtained by time-integrating the frequency offset over the pulse duration, weighted by the instantaneous intensity.

\begin{figure}
\begin{center}
\includegraphics[width=0.9\linewidth]{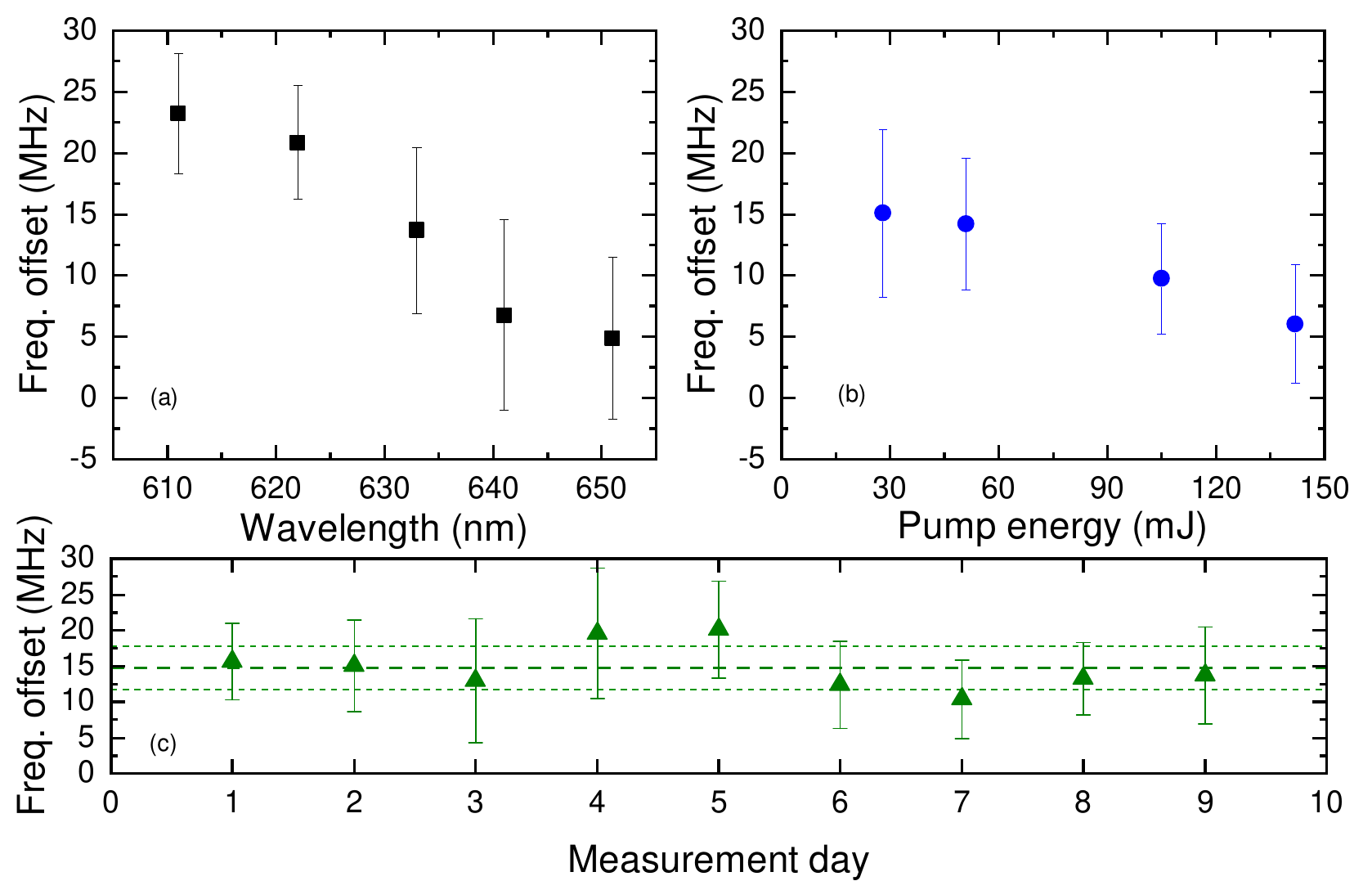}
\caption{\label{Chirp}
Result of frequency chirp measurements of the PDA laser system. a) Pulse-cw frequency offset Stokes beam from PDA (DCM in Methanol) at different wavelengths with 50 mJ pump energy.
b) Frequency offset of the PDA at 633 nm for different pump energies.
c) Daily variation of frequency offset of the PDA at 633 nm.
}
\end{center}
\end{figure}

A systematic study of frequency deviations between the cw-seed frequency and the PDA frequency was performed for various settings of the PDA.
Typical results for measured frequency offsets in the PDA, running on DCM in methanol, are displayed in Fig.~\ref{Chirp}. The chirp-offset is found to vary from 23 MHz at 610 nm (Stokes wavelength for \Tm) to 5 MHz at 650 nm (Stokes wavelength for HT) as shown in Fig.~\ref{Chirp}a, while also a dependence on the intensity was observed as shown in Fig.~\ref{Chirp}b. In addition, the effect of frequency chirp over the spatial beam profile of the Stokes beam was measured, and was observed to vary within 2 MHz.
During the CARS measurements the chirp analysis was performed after co-propagating the output of the pulsed laser with that of the AOM-shifted cw-laser output in a single mode fiber, so that this spatial effect was averaged out.
Systematic chirp-analysis measurements were performed
for the laser settings during every measurement day.
The frequency calibration of the PDA-laser was corrected for the chirp results during the CARS precision measurements, for which an estimated uncertainty of 5 MHz represents the standard deviation over different measurement days, as shown in Fig.~\ref{Chirp}c. 

The frequency of the 532 nm pump pulse, obtained from the injection-seeded Nd:YAG laser, was directly measured by a High Finesse WSU-30 (Toptica) wavemeter. The absolute calibration of the wavemeter was verified during each measurement run by measurements of I$_2$-hyperfine transitions in the range 600 to 650 nm, as well as measurement of a cw Ti:Sa laser at 718 nm locked to a frequency comb laser~\cite{Cheng2018}. 
These measurements result in a value for the typical drift of the wavemeter, determined at~0.3 MHz/hr.
In addition the effect of measuring cw laser vs. pulsed laser radiation was verified by measuring the frequency of the output of the cw ring laser seed frequency vs. the output of the PDA. As a result of these repeated procedures the uncertainty of the frequency of the pump pulse is estimated at 6 MHz.
In a recent study a similar model wavemeter (WSU-2) was used in measurements of several $^{40}$Ca$^{+}$ ion transitions and the calibrations methods led to an absolute accuracy of 5 MHz ~\cite{Stellmer2014}.

\subsection{ac-Stark shift}
\label{subsection:ac-Stark}

The ac-Stark effects occurring in CARS measurements of the hydrogen molecule have been investigated in detail~\cite{Rahn1980,Moosmuller1989,Dyer1991}. In our current setup, the ac-Stark shift is caused by both Stokes and pump pulses and can be approximated in terms of pulse energies $E_{P}, E_{S}$, areas at the beam waist $A_{P}, A_{S}$ and pulse durations (FWHM) $\tau_{P}, \tau_{S}$, and the ac-Stark shift can be expressed as:
\[\delta\nu = \kappa_P \frac{E_P}{A_P\tau_P} + \kappa_S \frac{E_S}{A_S\tau_S}.\] The constant $\kappa$, for both $P$ and $S$ depends on the difference of the ac-Stark shift in upper and lower states via:
\begin{equation} \label{eq:acstark}
\kappa = -\frac{2\pi}{hc} \left( (\alpha_{v'}-\alpha_{v''})
 + \frac{2}{3} \frac{J(J+1)-3m_J^2}{(2J+3)(2J-1)}(\gamma_{v'}-\gamma_{v''})\right)
\end{equation}
where $\alpha_{v'}$ and $\gamma_{v'}$ are the dynamic (frequency-dependent) isotropic polarizabilities and anisotropic polarizabilities at frequencies $v'$ and $v''$, respectively. The zero field transition frequency can be obtained by performing measurements at different Stokes and pump energies and extrapolation to zero. Since the frequencies used in CARS are far from the dipole-allowed transitions in molecular hydrogen, the polarizability term only weakly depends on the frequency in the optical range~\cite{Dyer1991} and the expression can be approximated as (with $\kappa_P=\kappa_S=\kappa$):
\[\delta\nu = \kappa\left(\frac{E_P}{A_P\tau_P} + \frac{E_S}{A_S\tau_S}\right).\] As the effective area and pulse width of the pump and Stokes beams are affected by spatial and temporal overlap between the two pulses, the value and the reproducibility of the shift coefficient in terms of pulse energy for pump and Stokes were measured extensively on the DT Q(1) transition~\cite{Lai2019}, yielding $\frac{\kappa_P}{A_P\tau_P}=-0.0064(7)$ \wn mJ$^{-1}$ and $\frac{\kappa_S}{A_S\tau_S}=-0.0006(2)$ \wn mJ$^{-1}$.
To find a balance between minimal ac-Stark shift and sufficient signal strength, the Stokes pulse energy was kept below 10 $\mu$J for the \Tm\ and DT measurements. In such conditions, the uncertainty of the ac-Stark shift from the Stokes beam, upon extrapolation to zero intensity, is estimated to be less than 1 MHz. Due to the weaker signal strengths obtained in the HT experiments, the Stokes energy was raised to 20 $\mu$J, while the pump energy was varied from 90 to 500 $\mu$J. Here the the pump energy was not increased further in view of saturation and asymmetry phenomena observed in the spectra, as discussed below.

\subsection{Pressure shift}

Collisional shift coefficients of Raman transitions in stable molecular hydrogen isotopologues are well studied in the large pressure range from tens of millibars to a few bars of total pressure~\cite{Rosasco1991,Rahn1990,Rosasco1989}. For the reported shift coefficients the presently measured CARS lines are estimated to undergo shifts below $\sim1$~MHz. In addition, the collisional shift of \Dm\ was investigated in the present study by pressure-dependent CARS measurements of pure \Dm\ in an identical gas cell. The shift coefficient is found to equal $0.06$ MHz/mbar under room temperature conditions.

For the \Tm\ measurements, the sample contains 93.4 \% of \Tm ~\cite{Schlosser2017}. The collisional shift is assumed to be similar to the self-collisional shift coefficient of \Hm\ and \Dm\, yielding a shift of about 0.3 MHz. From the reported result of HD~\cite{Rosasco1989}, the pressure shift in this heteronuclear species was found to be twice that of \Hm\ and \Dm, which was ascribed to the presence of a small permanent dipole moment and the lack of a selection rule prohibiting para-ortho conversion during rotationally inelastic collisions~\cite{Nazemi1983}. The DT and HT samples inevitably contain some amounts of \Dm\ and \Hm, respectively. We conclude that based on previous studies~\cite{Rosasco1991,Rahn1990,Rosasco1989}, and on present data for \Dm, the collisional shift in the present CARS experiments is well below 1 MHz.

Some of the CARS measurements on \Dm\ were performed from the DT gas sample cell, hence under conditions of a low-density plasma environment, undergoing constant production of about 10$^9$ electrons per second at an average kinetic energy of 5 keV and the same rate of ion production from the primary beta decays. In addition, secondary electrons and ions are produced from ionization of the neutral gas.
The transition frequencies of the \Dm\ CARS signals were not found to deviate from the signals from a different gas cell filled with 4 mbar of pure \Dm, without addition of radioactive species. We conclude that there occurs no measurable plasma shift on the transition frequencies as reported in Table~\ref{tab:Results}.

\subsection{Other effects}

There is underlying hyperfine structure in the \Tm, DT and HT transitions, where the hyperfine splitting span a range of less than a MHz. This hyperfine structure is similar to those in the stable molecular hydrogen species such as HD, which was shown to affect vibrational resonance lineshapes in the high-resolution investigations of Diouf et al. \cite{Diouf2019}. However, in view of the resolution of the present experiment this does not affect the determination of the transition frequency positions (hyperfine center-of-gravity).
Due to its closed-shell nature, the ground electronic state transitions in molecular hydrogen are insensitive to DC Stark shifts. Using the static polarizabilities from Ref.~\cite{Raj2019}, we estimate \cite{Dyer1991} an upper limit of well below 1~Hz for a DC field amplitude of 1~V/cm. Thus the effect due to the plasma generated from the $\beta$-decay would be higher, but this plasma effect was not observed in the \Dm\ measurements discussed in the previous section. 
Likewise, the Zeeman shifts of the hyperfine components are also very low, with estimated shifts using constants from Ref.~\cite{Ramsey1957} of well below $1$ Hz due to the earth's magnetic field.
In addition, for the Q transitions with the same upper $J'$ and lower $J''$ rotational quantum numbers and almost identical hyperfine constants for the $v=0$ and $v=1$ states, the shifts in the transition frequency effectively cancel out. 
We thus neglect the contributions of the hyperfine structure, DC Stark and Zeeman effects from the uncertainty budget.

\subsection{Frequency uncertainty in the CARS measurements}

Table~\ref{tab:unc} lists all the uncertainty contributions to the line center determination in the present CARS study. For \Tm\ Q($J=1-5$) and DT Q($J=0-5$), the total uncertainty is about 12 MHz. For the \Tm\ Q(0) and DT Q(6) lines the uncertainty is estimated at 20 MHz, due to interference effects, which will be discussed in the last section. For HT, the lowest pulse energy used in the measurement was about 2.5 times larger than that for \Tm\ and DT measurements, thus resulting in a larger uncertainty from the ac-Stark effects. In addition, the Doppler width for the HT lines is 1.4 times larger than that for DT. Such differences are also found in results from different measurement days. For HT, the total uncertainty is about $16$ MHz for the Q($J=0-3$) lines.

\begin{table}[b]
\begin{center}
\caption{\label{tab:unc}
Uncertainty budget for the CARS measurements.
Systematic and statistical contributions to the frequency uncertainties in the fundamental vibrational Raman Q($J<6$) transitions, excluding \Tm\ Q(0).
Values in units of $10^{-4}$ \wn\ and representing $1\sigma$.
}
\begin{tabular}{lS[table-format=2.1]S[table-format=2.1]}
Contribution & \multicolumn{1}{c}{\Tm\ / DT} & \multicolumn{1}{c}{HT}\\
\hline
Pump ($\omega_P$) calibration			& 2  	    & 2\\
Stokes ($\omega_S$) cw calibration		& 0.7  	    & 0.7\\
Stokes chirp correction		            & 1.7  	    & 1.7\\
ac-Stark analysis 				        & 2  	    & 3.3\\	
Collisional shift				        & 0.3	    & 0.3\\	
Plasma shift                            & 0.3       & 0.3\\
CARS interference shift                 & 1       & 1\\
Statistics 					            & 2.3  	    & 3\\	
\hline
Combined (1$\sigma$)				    & 4.2 	    & 5.3\\
\end{tabular}
\end{center}
\end{table}

\section{Results and Discussions}

\begin{figure*}
\begin{center}
\includegraphics[width=0.9\linewidth]{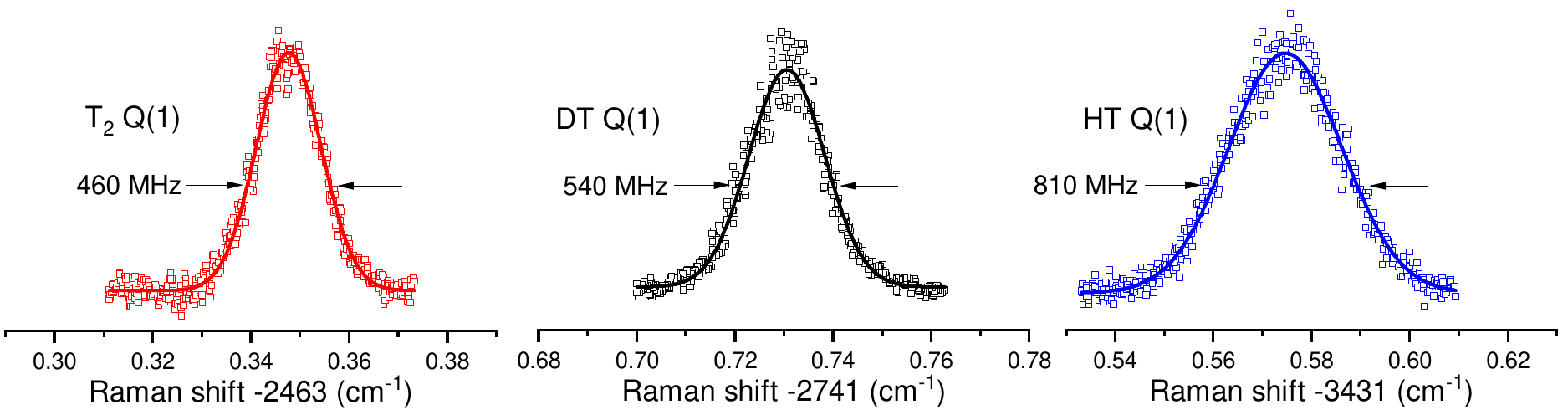}
\caption{\label{overview}
High resolution CARS spectra of the Q(1) lines in the fundamental vibrational band $(v=0\rightarrow 1)$ of T$_2$, DT, and HT.
}
\end{center}
\end{figure*}

Representative high-resolution CARS spectra of the ($v=0\rightarrow 1$) Q(1) transition for tritiated molecular hydrogen species \Tm, DT, and HT are displayed in Fig.~\ref{overview}. The line widths of the Q(1) transitions exhibit the expected Doppler widths with the broadest width for HT (lightest) and narrowest for \Tm\ (heaviest).
While the Q($J=0-5$) transitions in \Tm\ were previously reported in Ref.~\cite{Trivikram2018}, these lines have been remeasured in this study and the Q-branch extended up to $J=7$.
Moreover, the ($v=0\rightarrow 1$) S($J=0-3$) transitions, which are much weaker than the Q lines, were also investigated in the present work.
The DT ($v=0\rightarrow 1$) Q($J=0-5$) transitions were recently reported in Ref.~\cite{Lai2019} and extended here up to $J=7$.
The high-resolution measurements of the HT Q-branch transitions are presented here for the first time.
The transition frequencies of the Q-branch lines in the fundamental band splitting ($v=0\rightarrow 1$) of all tritiated species are listed in Table \ref{tab:Results}.

\subsection{HT, DT and T${}_2$ transitions}
The HT measurements include Q transitions from $J = 0-3$, where the signal strength involving higher $J$ states is limited due to decreasing population.
The HT partial pressure was somewhat lower than the targeted equilibrium partial pressure of 3.6 mbar due to issues in the HT gas filling procedure.
In addition, the lower transmission of the bandpass filter in the spectral region of the HT Q-branch lines, led to a weaker observed signal strength.
To compensate for the weaker signals higher pulse energies were employed to obtain sufficient SNR. This resulted, however, in a larger ac-Stark uncertainty.
These factors as well as the larger Doppler width lead to a larger total uncertainty in the HT transition energies than those of the DT and \Tm\ measurements.
The present Q-branch transition energies are 6.4 \wn\ higher than the values reported by Edwards et al.~\cite{Edwards1979}. These discrepancies were already pointed out in the subsequent studies of Chuang and Zare~\cite{Chuang1987} and Veirs and Rosenblatt~\cite{Veirs1987}.
The present HT Q-branch measurement results are consistent with and improve the measurement uncertainty of Veirs and Rosenblatt~\cite{Veirs1987} by a factor of 200.
Chuang and Zare~\cite{Chuang1987}  performed absorption measurements of weakly-allowed dipole P- and R-branch transitions in the (1,0) fundamental band, as well as the (4,0) and (5,0) overtone bands of HT.
When converting the transition frequencies of that study~\cite{Chuang1987} into fitted molecular constants, good agreement is found with the present results, which are more accurate by seventeen-fold.

\begin{table}[tp]
\caption{\label{tab:Results}
Fundamental vibrational $(v=0\rightarrow 1)$ splittings of the Q($J$) transitions in \Tm, DT, and HT. The weaker S($J=0-3$) transitions in \Tm\ are also listed.
The measured values appear in the second column while the theoretical values are listed in the third column, with uncertainties indicated within parentheses in units of the last digit in the indicated value.
The last column represents the difference ($\omega_{\rm{exp}}$ - $\omega_{\rm{calc}}$) with the combined uncertainty from experiment and calculation indicated. Experimental values include corrections from the interference phenomena affecting the CARS lineshape.
Lines indicated with asterisks (*) include uncertainty contributions from line profile asymmetry, while those with uncertainties of $5 \times 10^{-3}$ \wn\ are dominated by ac-Stark broadening.
All values in units of \wn.}
\begin{tabular}{llll}
Line	& \multicolumn{1}{c}{Experiment}	& \multicolumn{1}{c}{Calculation}	& \multicolumn{1}{c}{Difference}\\
\hline
 \multicolumn{4}{c}{\textbf{T$_2$}}   \\
Q(0)	& 2\,464.503\,94\,(67)$^*$ & 2\,464.504\,15\,(6)& $-$0.000\,21\,(67)\\ 
Q(1)	& 2\,463.348\,17\,(42) & 2\,463.348\,36\,(6)& $-$0.000\,19\,(42)\\
Q(2)	& 2\,461.039\,17\,(42) & 2\,461.039\,17\,(6)& $+$0.000\,00\,(42)\\
Q(3)	& 2\,457.581\,35\,(42) & 2\,457.581\,37\,(6)& $-$0.000\,02\,(42)\\
Q(4)	& 2\,452.982\,33\,(42) & 2\,452.982\,11\,(6)& $+$0.000\,22\,(42)\\
Q(5)	& 2\,447.250\,61\,(42) & 2\,447.250\,85\,(6)& $-$0.000\,25\,(42)\\
Q(6)	& 2\,440.397\,(5) & 2\,440.399\,34\,(6)& $-$0.002\,(5)\\
Q(7)	& 2\,432.442\,(5) & 2\,432.441\,52\,(6)& $+$0.000\,(5)\\
S(0)	& 2\,581.114\,(5) & 2\,581.105\,22\,(6)& $+$0.009\,(5)\\
S(1)	& 2\,657.281\,(5) & 2\,657.282\,90\,(6)& $-$0.002\,(5)\\
S(2)	& 2\,731.716\,(5) & 2\,731.710\,84\,(6)& $+$0.006\,(5)\\
S(3)	& 2\,804.164\,(5) & 2\,804.164\,21\,(6)& $-$0.000\,(5)\\
 \multicolumn{4}{c}{\textbf{DT}} \\
Q(0)	& 2\,743.341\,60\,(42) & 2\,743.341\,74\,(11)& $-$0.000\,14\,(43)\\
Q(1)	& 2\,741.732\,04\,(39) & 2\,741.732\,10\,(11)& $-$0.000\,06\,(40)\\
Q(2)	& 2\,738.516\,62\,(42) & 2\,738.516\,97\,(11)& $-$0.000\,35\,(43)\\
Q(3)	& 2\,733.704\,79\,(42) & 2\,733.704\,66\,(11)& $+$0.000\,13\,(43)\\
Q(4)	& 2\,727.307\,45\,(42) & 2\,727.307\,55\,(11)& $-$0.000\,10\,(43)\\
Q(5)	& 2\,719.342\,21\,(42) & 2\,719.342\,02\,(11)& $+$0.000\,19\,(43)\\
Q(6)	& 2\,709.828\,35\,(67)$^*$ & 2\,709.828\,32\,(11)& $+$0.000\,03\,(68)\\ 
Q(7)	& 2\,698.787\,(5) & 2\,698.790\,48\,(11) & $-$0.003\,(5)\\
 \multicolumn{4}{c}{\textbf{HT}} \\
Q(0)	& 3\,434.812\,48\,(53) & 3\,434.813\,33\,(44)& $-$0.000\,85\,(69)\\
Q(1)	& 3\,431.575\,09\,(53) & 3\,431.575\,53\,(44)& $-$0.000\,44\,(69)\\
Q(2)	& 3\,425.112\,65\,(53) & 3\,425.113\,24\,(44)& $-$0.000\,59\,(69)\\
Q(3)	& 3\,415.452\,58\,(53) & 3\,415.452\,98\,(44)& $-$0.000\,40\,(69)\\
\hline
\end{tabular}
\end{table}

Measurements of the DT Q-branch were reported in our recent work~\cite{Lai2019} and are extended here to include Q(6) and Q(7) lines. The uncertainty for the much weaker Q(7) line is substantially larger than those for $J<5$, caused mainly by the asymmetric line profile, which is discussed below.
A comparison of the present DT Q-branch transition frequencies with the Q($J=0-4$) values reported by Edwards et al.~\cite{Edwards1979} shows differences of some 0.15 \wn, hence much larger than their claimed uncertainty of 0.005 \wn.~\cite{Edwards1979}
\, Veirs and Rosenblatt~\cite{Veirs1987} already had cast doubt on the accuracy claimed by Edwards et al.~\cite{Edwards1979} given the discrepancies in the analogous HT and \Tm\ measurements.
The DT Q-branch measurements presented here are consistent with those of Veirs and Rosenblatt~\cite{Veirs1987} and improve the measurement uncertainty by a factor of more than 200 for the Q($J=2-7$) transitions.

\begin{figure}
\begin{center}
\includegraphics[width=0.75\linewidth]{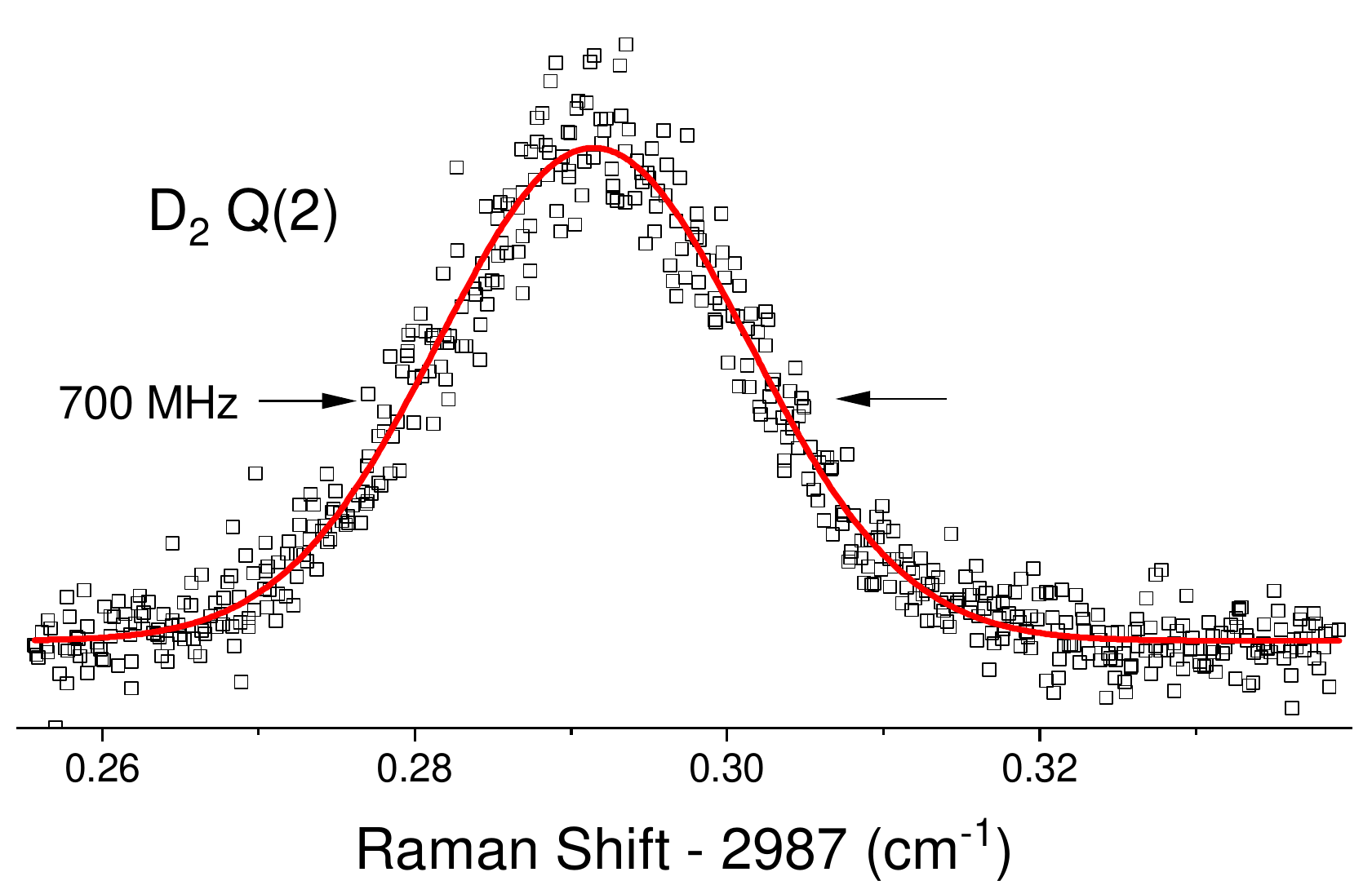}
\caption{\label{D2_cw-stokes}
Quasi-cw CARS spectra of the Q(2) line in the fundamental band of \Dm\ recorded at 8 mbar using continuous wave Stokes radiation at 80 mW and a 8-ns pump pulse with 0.47 mJ pulse energy.
}
\end{center}
\end{figure}

The present \Tm\ Q($J = 0 - 5$) values were obtained after implementing improvements in the setup used in Ref.~\cite{Trivikram2018}.
Comparing the \Tm\ results listed in Table~\ref{tab:Results} with the values reported in Ref.~\cite{Trivikram2018}, reveals an offset of $<0.001$ \wn\ for the Q(0), Q(1), and Q(3) transitions while showing consistent values for Q(2), Q(4) and Q(5).
We attribute these discrepancies to several difficulties in the measurements of \Tm\ in the previous study.
For example, the cw-seed instability of the pump laser for both CARS laser beams resulted in substantial time jitter leading to poor SNR, especially for weaker transitions.
To compensate for such signal loss, the Stokes and the pump pulse energies used during the previous experiment were much higher than in the present study.
The cross-sectional areas of both laser beams are now more accurately determined using two methods, a CCD camera beam profiler system and knife-edge method, leading to consistent values.
Compared to the estimates of laser intensities in the previous study,~\cite{Trivikram2018} these could be off by a factor of three, which is important in the ac-Stark extrapolation studies.
A set of measurements, carried out during a day, is based on an extrapolation relying only on relative intensities, assuming little alignment drift, and should lead to robust results.
This becomes an issue when combining results of different days where any misalignment may lead to a slightly different overlap of the pump and Stokes laser beams, yielding a different effective area for production of CARS signal. 
Moreover, in the previous study the ac-Stark analysis for all transitions except Q(1) was performed by altering the pump beam intensity only, and thus the ac-Stark shift contribution due to differences in the Stokes beam overlap could be another source for the offset. These shortcomings were repaired in the present study leading to more reliable results.

The frequency calibration of the Stokes laser is also improved in the present study.
Some transitions exhibit a more than 0.15 \wn\ separation from the calibration line, represented by a particular I$_2$ hyperfine component.
Since such frequency span covers more than 30 markers of the reference etalon, length stabilization is now ensured over the whole recording time by locking to a stabilized HeNe-laser. Improved frequency calibration of the free spectral range (FSR=150.67(2) MHz) of the reference cavity is also applied. 
The issues discussed here could have led to an underestimate of the uncertainties in the previous study~\cite{Trivikram2018}.
In addition, the present measurements of \Tm\ lines in the improved setup allowed for recordings at lower pulse energies than in Ref.~\cite{Trivikram2018}, resulting in a reduced uncertainty of ac-Stark shift described in the previous section.
These improvements have reduced systematic shifts, while at the same time have achieved better sensitivity and enhanced SNR of the recordings.
These in turn enabled extending the measurements to weaker Q(6) and Q(7) transitions, as well as the \Tm\ S-branch with the transition frequencies listed in Table~\ref{tab:Results}.
The \Tm\ Q($J=0-5$) results obtained in our new measurements represent a nearly 200-fold improvement compared with results from spontaneous Raman spectroscopy by Veirs and Rosenblatt~\cite{Veirs1987}.

\begin{figure*}
\begin{center}
\includegraphics[width=\linewidth]{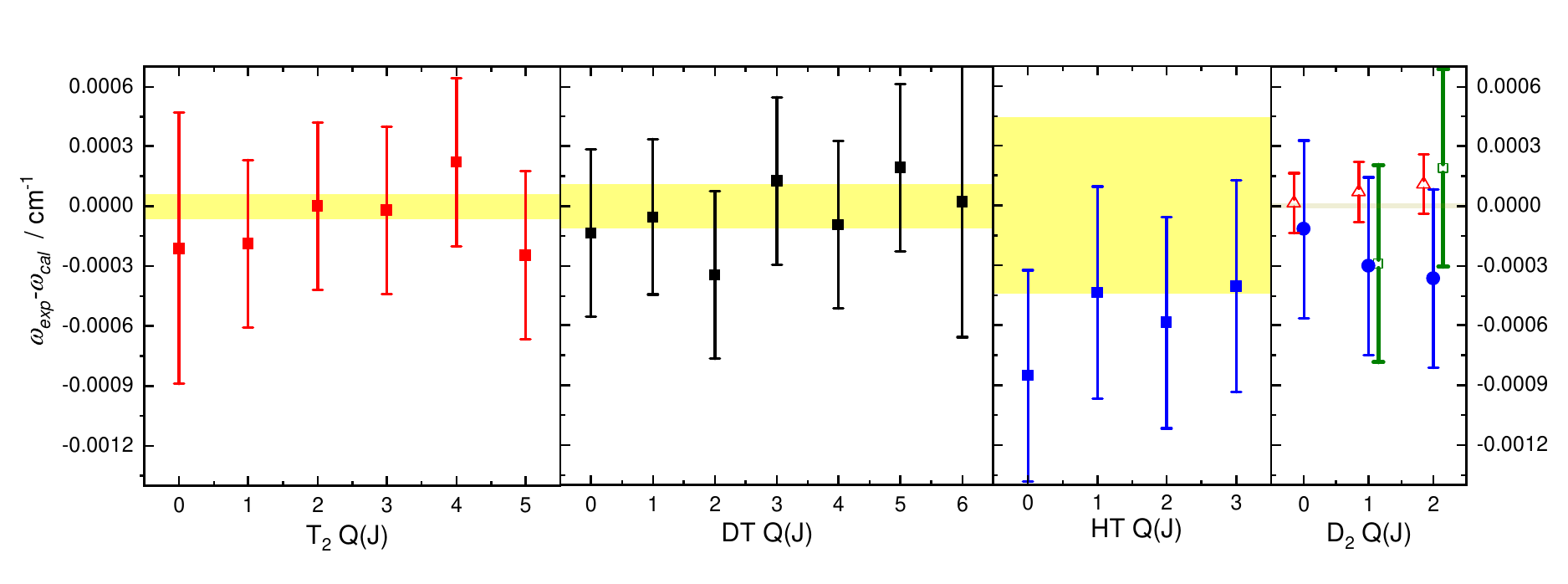}
\caption{\label{Summary}
Comparison of the experimental and calculated fundamental vibrational splittings of all tritiated species.
The error bars on the data points indicate the measurement uncertainty while the (yellow) shaded regions indicate the uncertainty of the calculations.
The rightmost trace shows results of CARS spectra for \Dm, measured in a molecular beam expansion (red triangle), from a quasi-cw CARS measurement (green square), and measurements with pulsed pump and Stokes laser beams (blue circle) as used for the tritiated species.
The \Dm\ points are slightly displaced horizontally for clarity.}
\end{center}
\end{figure*}

\subsection{Benchmark measurements on D${}_2$}

CARS measurements on pure \Dm\ were performed in an identical gas cell focusing on studies of systematic effects, such as the validation of pressure shift effects.
This allowed for the comparison of \Dm\ Q(0), Q(1) and Q(2) of the fundamental vibrational splitting with previous precision studies using molecular beams and a deep-UV REMPI scheme~\cite{Dickenson2013, Niu2014}.

In addition, quasi-cw CARS measurements were performed on \Dm\ Q(1) and Q(2) lines, using a cw Stokes beam, in combination with a pulsed pump beam. A quasi-cw CARS spectrum of the \Dm\ Q(2) transition is shown in Fig.~\ref{D2_cw-stokes}. Such experiments bear the advantage that frequency chirp effects in the Stokes beam are absent. With 80 mW of cw Stokes radiation and a 20 $\mu$m beam waist, it was possible to detect the CARS signal using a pump energy of as low as 300 $\mu$J in a cell containing 8 mbar of \Dm\ gas. The accuracy of these quasi-cw CARS measurements is estimated to be 15 MHz, limited by the ac-Stark extrapolation.

Results of a number of test experiments performed on \Dm\ are included in the rightmost panel of Fig.~\ref{Summary}. The results of these studies using pulsed pump and Stokes radiation for Q($J=0-2$), as well as using pulsed pump and cw Stokes beams for Q(1) and Q(2) are in good agreement with the precision molecular beam experiments in Refs.~\cite{Dickenson2013, Niu2014}.
These test experiments on \Dm\ verify the assessment of pressure shifts in the present CARS study. 
In addition, the use of chirp-free Stokes radiation in the quasi-cw CARS measurements on D$_2$ validates the chirp correction procedures implemented for the majority of data recorded with the PDA-laser system.

\subsection{Comparison with calculations}

Ab initio calculations, in the framework of NAPT, on level energies in the ground electronic state, are described in Refs.~\cite{Lai2019,Komasa2019} for all tritiated species. Rovibrational level energies and transition frequencies for all molecular hydrogen isotopologues in the ground electronic state can now be accessed through a publicly available program~\cite{SPECTRE}.
The differences between experimental results and calculated values, expressed as $\omega_\mathrm{exp}-\omega_\mathrm{calc}$, are plotted in Fig.~\ref{Summary} and listed in Table ~\ref{tab:Results}.

In Fig.~\ref{Summary}, the experimental values are represented by data points with error bars attached, while a yellow-shaded region represents the uncertainty of the calculation. Very good agreement between measurement and theory is found for all tritiated species, as can be read from Fig.~\ref{Summary}.
For the tritium-bearing species, the major contribution to the theoretical uncertainty derives from the non-relativistic energy, $E^{(2)}$, which is about $\sim6\times10^{-5}$ \wn\ for the heaviest \Tm\ species, and about $\sim 4 \times10^{-4}$ \wn\ for HT, owing to its lighter mass~\cite{Lai2019}. 
Hence, the results from the calculations are more accurate than from experiments for \Tm\ and DT, while for the case of HT a similar accuracy is obtained, due to the less accurate $E^{(2)}$ term.

For the stable species \Hm, HD and \Dm, the $E^{(2)}$ terms have been calculated directly in a 4-particle variational approach (i.e. without invoking the Born-Oppenheimer approximation and corrections, hence outside the NAPT-framework), achieving $10^{-7}$ \wn\ uncertainty levels~\cite{Komasa2019}.
Therefore, the frequencies for the \Dm\ species are extremely accurate ($1.2\times10^{-5}$ \wn)~\cite{Komasa2019}, as represented by the narrow yellow strip in the rightmost panel of Fig.~\ref{Summary}.

A similar calculation of improved accuracy might in future be performed for the tritiated species, thereby reducing the total uncertainty to the $10^{-5}$ \wn\ level, limited by higher-order $E^{(5)}$ and  $E^{(6)}$ terms.
However, the current theoretical uncertainty of the NAPT-framework is sufficient for a comparison with experimental values, with uncertainties on the order of a few $10^{-4}$ \wn\ for the tritiated molecular hydrogen species.

\section{Lineshape analysis}

For most lines observed at good signal-to-noise ratios presented in this study, the lineshapes were found to be symmetric. These resonances were fitted with Voigt profiles, dominated by a Gaussian component with Doppler and instrumental contributions.
At sufficiently high intensities for strong transitions, saturation dips are observed and were in fact exploited for \Tm\ spectroscopy in Ref.~\cite{Trivikram2018} for the ac-Stark extrapolation. We discuss this saturation effect in this section and its consequence for extracting the line position from the profile.
For the weaker transitions such as those in the S-branch of \Tm, some asymmetry in the lineshape is observed. 
We discuss several contributions that result in the asymmetric profiles including the ac-Stark effect, non-resonant background and interference effects occurring in the non-linear optical CARS process.

\subsection{Saturation CARS spectroscopy}

The accuracy of the current CARS measurements is mainly limited by the Doppler-broadened linewidth. In view of testing future improvements of QED-calculations beyond the NAPT-framework experimental uncertainties at the level of $< 10^{-5}$ \wn\ would be desired. Such reduced measurement uncertainties might be achievable via saturated CARS spectroscopy, where a narrow Doppler-free Lamb dip is detected. Saturated CARS profiles were explored by Lucht and Farrow~\cite{Lucht1989} who proposed a theoretical description via a density matrix formalism. In that study, a mechanism for the saturation effect was invoked either via strong population pumping or by spectral interference between different velocity classes. The population hole-burning effect in saturation spectroscopy gives rise to typical Lamb dips at the resonance positions. The spectral interference effect, discussed in detail below, results in the cancellation of the real part of the third order susceptibility of a particular velocity class $\overrightarrow{u}$ with that of the opposite direction $-\overrightarrow{u}$. Such cancellation is also most pronounced at the resonance position, which then also contributes to saturation of the CARS signal.
In a different approach Owyoung and Esherick~\cite{Owyoung1980} have demonstrated saturation in stimulated Raman loss spectroscopy of \Dm\ in a fashion of typical pump-probe spectroscopies, using a three and four-beam counterpropagating configuration. 
The observed saturation dip exhibited a width of 110 MHz, limited by laser bandwidth~\cite{Owyoung1980}.
This suggests that with a narrower laser bandwidth and improved frequency calibration procedures, sub-MHz accuracies could potentially be achieved for tritiated species.

\begin{figure*}
\begin{center}
\includegraphics[width=0.9\linewidth]{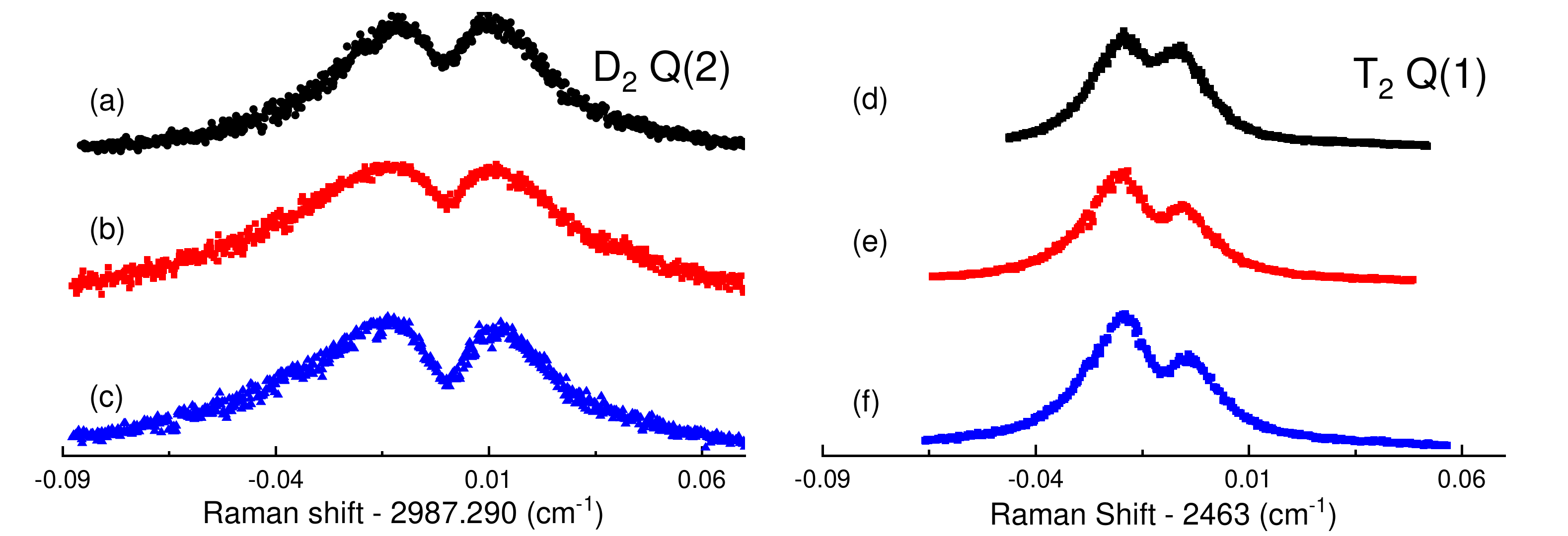}
\caption{\label{Saturation_Power_dependence_T2_D2}
Observed spectral profiles for CARS signals exhibiting saturation Lamb dips for two lines recorded at different pump pulse energies.  Specific conditions: 
\Dm\ Q(2) with Stokes pulse energy 0.74 mJ and pump energy:
a) 0.58 mJ;
b) 1.2 mJ;
c) 1.7 mJ;
and \Tm\ Q(1) with Stokes pulse energy 1 mJ and pump energy:
d) 1.5 mJ;
e) 1.75 mJ;
f) 2 mJ.
}
\end{center}
\end{figure*}

We performed systematic studies of saturation CARS on the \Dm\ Q(2) and \Tm\ Q(1) transitions.
The \Dm\ Q(2) measurements were performed using a gas cell containing 4 mbar of pure \Dm.
A clear saturation feature appears when the pump and Stokes intensity product is $I_P\cdot I_S\sim5$ GW$^2$/cm$^{4}$. Fig.~\ref{Saturation_Power_dependence_T2_D2} shows recordings at various pump pulse energies.
As shown, the width of the Lamb dip in \Dm\ is about 300 MHz, while the Doppler-limited profile is more than 1.2 GHz wide, with width contributions from the instrument function and power broadening.
The spectra were fitted with two Gaussian function components, with opposite amplitude signs to represent the saturation Lamb dip and the Doppler-broadened profile.
It turned out that the linear ac-Stark extrapolation of the Lamb dip feature and that of the Doppler-broadened envelope do not produce consistent values for the zero-field transition frequency.
The Lamb-dip positions plotted against intensity shows a slope that is a factor of five smaller than expected from the unsaturated Doppler-broadened profile.
We suspect that this stems from 1) the spatial intensity distribution in the sample volume, which is difficult to integrate over, especially for the Lamb dips exhibiting a different nonlinear response to intensity than that of the Doppler-broadened envelope; and 2) the occurrence of CARS cross-interference effect~\cite{Bombach1990} on saturation. 

Saturated spectra of the \Tm\ Q(1) transition, recorded at similar alignment and intensity conditions as for \Dm\ Q(2), are shown in Fig.~\ref{Saturation_Power_dependence_T2_D2}.
The \Tm\ Q(1) saturation dip is observed to be shifted stronger at the higher-frequency flank with respect to the center of the Doppler-broadened profile at increasing pump energy.
The resulting asymmetric profile appears to be qualitatively different from the case of \Dm, where the Lamb dip center coincides with the Doppler envelope center at lower intensities.
Trivikram et al.~\cite{Trivikram2016} studied a related phenomenon on the electronic transitions of \Hm, where the line shift no longer follows a linear trend at sufficiently high intensities as the transitions exhibit skewed line profiles.
This presents a serious difficulty in the ac-Stark extrapolation, as high intensity measurements cannot be used in a linear extrapolation to the ac-Stark free value.
The intensity values reported in the previous \Tm\ study\cite{Trivikram2018} was probably overestimated thus staying in the linear regime for the ac-Stark shift.
In the present study, the line positions of the \Tm\ Lamb dip positions in Fig.~\ref{Saturation_Power_dependence_T2_D2}(d)-(f) already showed a much smaller power-dependence slope compared to measurements for unsaturated profiles at lower intensity.

A more accurate treatment of these phenomena needs to address the asymmetric ac-Stark broadening due to spatial and temporal intensity variations, such as those discussed in Refs.~\cite{Trivikram2016,Li1985}, which is not straightforward in the case of CARS.
In principle the Lamb-dips should produce more accurate results in determining transition frequencies, however, more refined studies are required to obtain a quantitative understanding of the saturation effect in CARS and to extract the most accurate line positions from the narrower resonance features.

\subsection{Asymmetric profile of weak transitions}

The CARS measurements were extended to transitions involving higher rotational states up to $J = 7$ for \Tm\ and DT and also to the weaker S-branch of  \Tm, with $\Delta J = +2$. The lineshapes for these weaker lines were observed to give rise to asymmetric profiles. These effects are partly caused by the required high intensity in the measurements to compensate for the weakness of transition probabilities, either from having a smaller Raman cross-section or a lower population.
The weak transitions are barely detectable with 1 mJ pump and Stokes energy and show different degrees of asymmetry in their spectral profile.
Nearly all these weak transitions show different tailing behavior in the flanks below and above the resonance and, despite the higher pulse energy used, no saturation effect at resonance is apparent.
As a consequence the accuracy for which the Q($J = 6,7$) transition frequencies in \Tm\ and DT, as well as \Tm\ S($J=0-3$) lines can be determined is much lower than for the strong CARS lines (the transition frequencies of these weak lines are included Table~\ref{tab:Results}).
We discuss three possible effects contributing to the asymmetry: 1) spatial beam profile and time evolution of pulses; 2) ac-Stark splitting of the magnetic sublevels, and; 3) resonant cross interference between lines and interference with the non-resonant background~\cite{Tolles1977}.

The first two effects were discussed by Moosm\"uller et al.~\cite{Moosmuller1989}, studying the spectral profile from coherent Stokes Raman scattering (CSRS) of N$_2$ exhibiting the effects of spatial and temporal variation of intensity.
Similarly as for the CARS profiles investigated here, molecules at different positions in the interaction volume and throughout the time evolution of tightly focused pump and Stokes pulses will have varying contributions to the overall spectrum, because of the spatial and temporal variation of the intensities.
It is expected that signal contributions near the focal point, which is highly red-shifted by ac-Stark effect in this case, are the strongest.

Apart from the ac-Stark shift due to the spatial and temporal intensity distribution, the ac-Stark induced splitting of the $m_J$ sublevels, which are unresolved in the measurements, will lead to an asymmetry of the spectral profile.
From Eq.~(\ref{eq:acstark}), the second term refers to the ac-Stark splitting for different $m_J$, where higher $\vert m_J\vert$ values experience less shift and overall produce a broadening of the spectrum. Such phenomena of ac-Stark splittings in CARS were quantitatively investigated previously~\cite{Dyer1991,Farrow1982}. Here we address such effects phenomenologically in the line fitting of asymmetric lines (see below).

\begin{figure*}
\begin{center}
\includegraphics[width=0.9\linewidth]{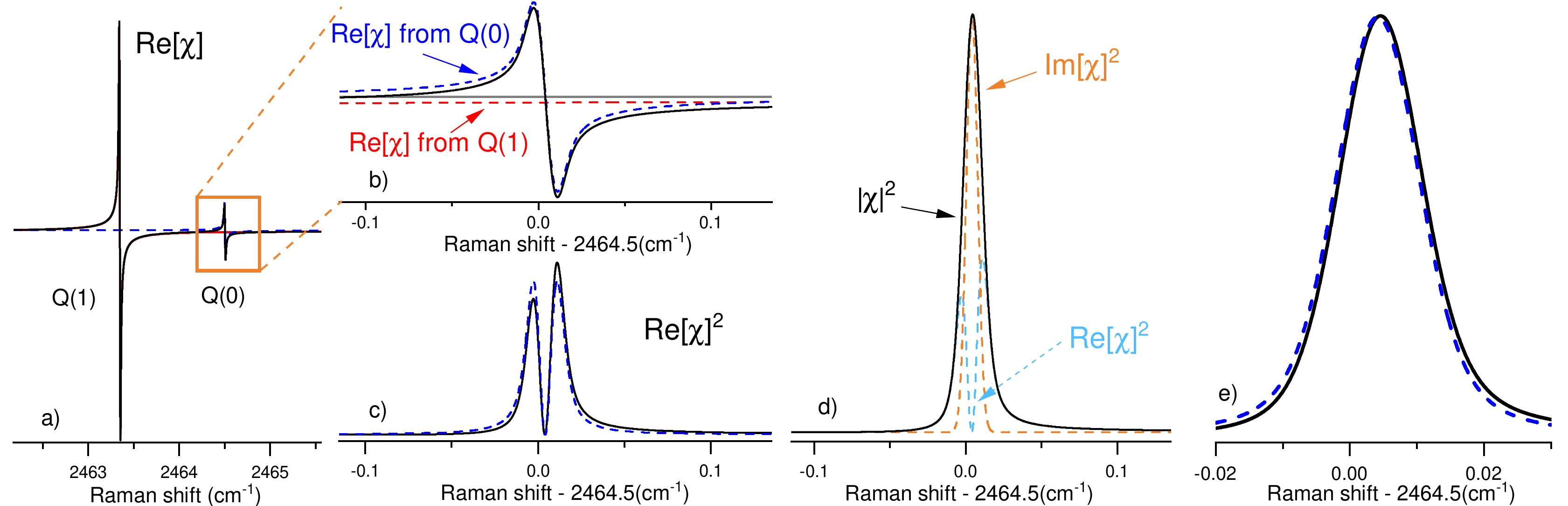}
\caption{\label{nr_simulation_T2}
Simulation of the CARS spectrum of the Q(0) line of \Tm, focusing on the effect of CARS spectral cross-interference.
The real and imaginary contributions of the Q(0) line are drawn in subfigures a) to d). See text for extensive discussion.
e) The line profile of the Q(0) line with the cross-interference effect is shown as the solid curve, while the dashed curve represents the case when the effect of interference is neglected.
The cross-interference causes a shift of +13 MHz of the Q(0) peak position.
}
\end{center}
\end{figure*}

Another well-known artifact in CARS spectra is the spectral interference in the third order susceptibility $\chi^{(3)}$.
The main contribution to the third order non-resonant susceptibility at 532 nm for the pump and 683 nm for the Stokes wavelengths is from the electronic polarizability, with only a smaller contribution due to molecular vibration and rotation~\cite{Lundeen1983}.
In Eq.~(\ref{eq:Ias}), the third-order susceptibility $\chi^{(3)}=\chi_{NR}+\chi_R $ comprises of a resonant and non-resonant part giving rise to interference effects in the CARS signal, $I_{AS}\propto |\chi_{NR}+\chi_R |^2$ (where the superscript $^{(3)}$ is dropped).
The real and positive $\chi_{NR}$, from far off-resonance contributions, will interfere with the real part of $\chi_R$ and distort the spectrum resulting in a dispersive-like signal.

In the following, we first discuss a numerical study to simulate the effects of the spectral interference effect on the lineshape that eventually shifts the peak position of the resulting profiles. Such shifts are included in the frequency determinations as listed in Table~\ref{tab:Results}. Thereafter, a fitting procedure is described using an asymmetric profile applied to the experimental spectra to assess shifts in the peak position.
In the latter, it remains difficult to decouple the separate contributions of the ac-Stark effect and spectral interference, giving at best an estimate of possible systematic frequency shifts due to the asymmetric profile. For these cases the uncertainties in the frequency values as listed in Table~\ref{tab:Results} were appropriately enlarged.

\subsection{Lineshape simulation for interference effects}

Following the analyses in Refs.~\cite{Druet1979,Rosasco1985} simulations were performed to account for the interference effects in the CARS lineshape that is proportional to the square of the third-order nonlinear susceptibility $|\chi|^2$:
\begin{equation} \label{eq:CARSchi3a}
|\chi|^2 = \left\vert\chi_{NR} +  \sum_J a_0(J) \int_{-\infty}^\infty \frac{\frac{1}{\sqrt{\pi}\, u}e^{-v_z^2/u^2}dv_z}{\omega_0(J)-(\omega_P - \omega_S) -i\Gamma - \frac{1}{2\pi}(k_P - k_S)v_z} \right\vert^{2},
\end{equation}
where $\omega_0(J)$ is the resonance frequency of a particular transition, $\Gamma$ is the relaxation rate, $(k_P - k_S)v_z/(2\pi c)$ is the Doppler shift, and $u$ is the most probable speed $\sqrt{2k_B T/M}$ for temperature $T$.
The integration over the Maxwell-Boltzmann distribution for all velocities $v_z$ along the CARS beam propagation direction can be performed and the equation recast using a Faddeeva function $w(\zeta)$
\begin{equation} \label{eq:CARSchi3}
|\chi|^2 = \left\vert\chi_{NR} - i \, \sum_J a_0(J) \, \frac{\sqrt{\pi}c}{\omega_R u} \, w\left(\frac{-\omega_0+\omega_R+i\Gamma}{\omega_R\frac{u}{c}}\right)\right\vert^{2}
\end{equation}
where $\omega_R=(\omega_P - \omega_S)$.
The ($v=1, J$) states are extremely long-lived, hence the Lorentzian width $\Gamma$ is set to zero in the present low-pressure and Doppler-limited case.
The $a_0(J)$ coefficients for the Q-branch transitions,
\begin{equation} \label{eq:chir}
a_0(J) = \frac{Np(J)}{h} \left(\alpha^1_0(J)^2+\frac{4}{45}b^J_J \gamma^1_0(J)^2\right),
\end{equation}
depend on the population number density $Np(J)$ for the rotational ground states $J$ and $b^J_J$ is the Placzek-Teller coefficient.
The isotropic $\alpha^1_0$ and anisotropic $\gamma^1_0$ polarizabilities between $v=1$ and $v=0$ are estimated from that of \Hm, HD and \Dm\ from Ref.~\cite{Raj2018}, since no values for the tritiated species are available.
For $J=0$, $\alpha^1_0(0)= 0.61$ a.u. and $\gamma^1_0(0)=0.50$ a.u. and these values do not vary significantly for other $J$ states.

There are two contributions to the interference effect in CARS. Firstly, the summation over all populated states gives rise to cross-interferences $\chi_{R}$ between all lines in the $\nu=0\rightarrow 1$ fundamental band, and secondly a nonresonant term $\chi_{NR}$ that represents the contribution from the far-lying electronic resonances. 
The $\chi_{NR}$ value is estimated from the \Hm\ value of $6.7 \times 10^{-18}$ cm$^3$/erg\,amagat given in Ref.~\cite{Hahn1995}.
For the 2.5 mbar \Tm\ sample at 298 K, this value converts to $1.5 \times 10^{-20}$ cm$^3$/erg or $1.5 \times 10^{-20}$ esu, while for the DT sample with 12.7 mbar total pressure a value of $7.6 \times 10^{-20}$ esu is estimated. 
The presence of \Dm\ in the DT sample results in additional cross-resonance contribution from the \Dm\ Q-branch (Raman shift at $\sim$2991 \wn) on the DT Q-branch (Raman shift at $\sim$2743 \wn) which is estimated to be $1.8 \times 10^{-20}$ esu.

Fig.~\ref{nr_simulation_T2} illustrates the effect of CARS interference on the \Tm\ Q(0) line, for which the dominant interference contribution is from the neighboring \Tm\ Q(1) transition. Since Q(1) is the most intense and the nearest lying resonance, it produces the strongest effect as follows from Eq.~(\ref{eq:CARSchi3}).
The real part $\mathrm{Re}[\chi]$ is plotted in Fig.~\ref{nr_simulation_T2}b showing the spectral region that spans \Tm\ Q(0), with the grey horizontal line indicating zero amplitude.
The dashed blue dispersive line belongs to Q(0) while the dashed red line below zero represents the contribution of the Q(1) line, yielding the solid blue dispersive line as the sum.
At the Q(0) resonance position, the nonresonant contribution $\chi_{NR}$ is more than a hundred times smaller than the resonant $\chi_{R}$ contribution from Q(1).
It is noted that this holds for the present specific case, given the rather low pressures in the sample yielding a small $\chi_{R}$.
The square of the real part of $\chi$ is plotted in Fig.~\ref{nr_simulation_T2}c, with the dashed curve indicating the pure contribution from Q(0) and the solid curve from the summation.
Fig.~\ref{nr_simulation_T2}d, shows both contributions of the real $\mathrm{Re}[\chi]$ and imaginary parts $\mathrm{Im}[\chi]$ of the third-order susceptibility as well as the square $|\chi|^2$.
Finally, Fig.~\ref{nr_simulation_T2}e shows the line profile of the \Tm\ Q(0) line. The solid line includes the cross-interference effect, while the dashed curve represents the case when the effect of interference is neglected.
This numerical example demonstrates how the large CARS susceptibility contribution from \Tm\ Q(1), due to the factor of 7 higher population in $J=1$ compared to that of $J=0$ (under room-temperature conditions), results in a large interference with the nearby Q(0) transition leading to a +13 MHz shift in Q(0).
The positive sign of the shift in \Tm\ Q(0) is due to the negative contribution of the real part of the $\chi$ susceptibility from the Q(1) transition (see Fig.~\ref{nr_simulation_T2}b).

\begin{figure}
\begin{center}
\includegraphics[width=0.9\linewidth]{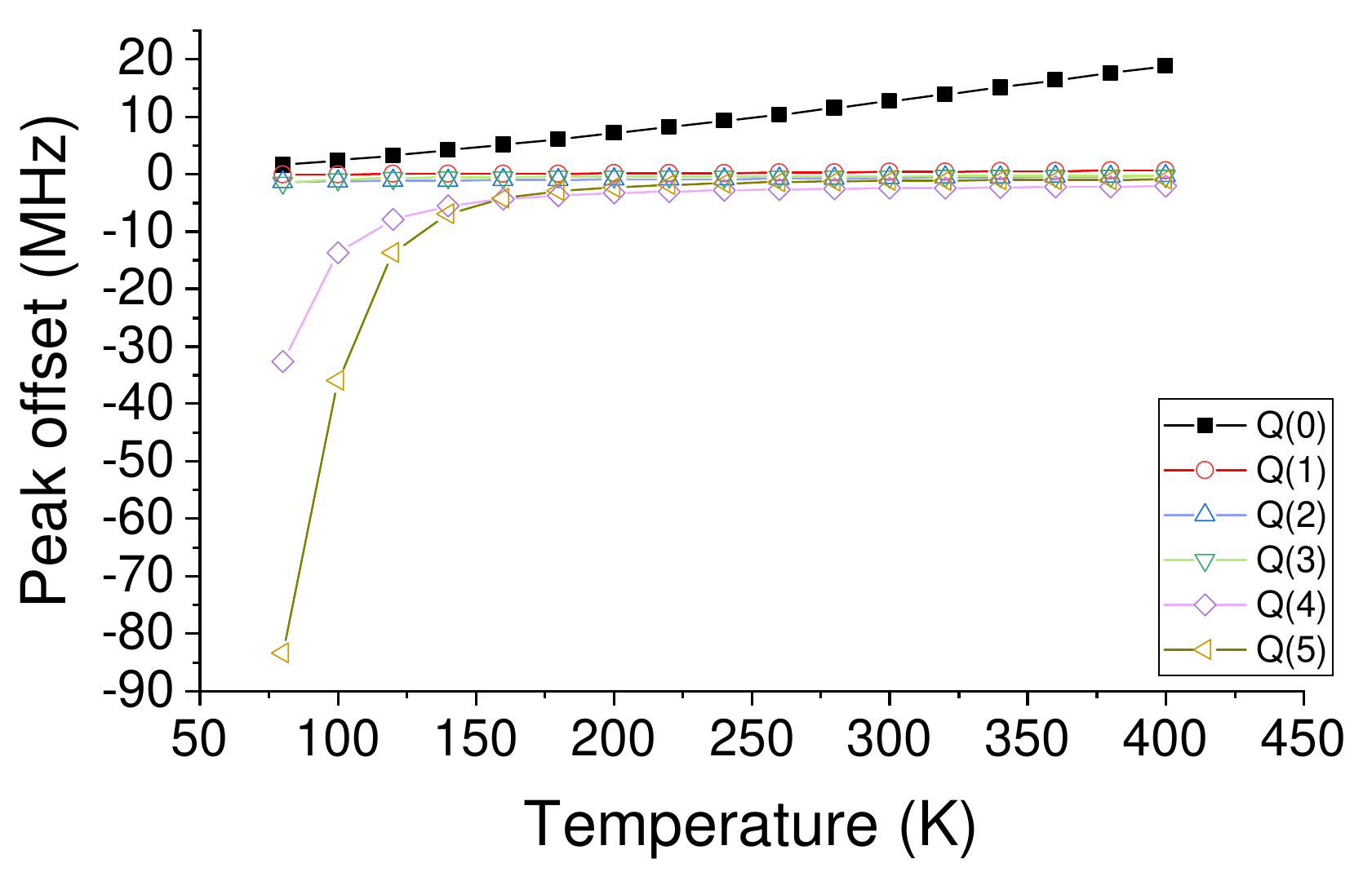}
\caption{\label{T2_peak_offset}
Simulated shifts of peak positions of the \Tm\ Q-branch transitions due to spectral cross-interference as a function of temperature.}
\end{center}
\end{figure}

Systematic shifts of the line centers due to spectral interference were simulated for all \Tm\ Q-branch transitions as a function of temperature, and results are shown in Fig.~\ref{T2_peak_offset}.
The Q(0) line is blue-shifted due to cross-interference with other Q($J\neq 0$) lines.
For increasing temperature there are two effects that enlarge the cross-interference shift on the Q(0) line: the population ratio between $J = 1$ and $J = 0$ grows and approaches 9:1 while also the larger Doppler width leads to a stronger interference experienced by Q(0).
The sharp drop in the peak position offset of the Q(4) and Q(5) lines below 200 K is due to the drastic decrease in population of the $J=4,5$ states.
At around 300 K where the measurements were performed, systematic shifts of the line centers for the \Tm\ Q($J=1-5$) lines were found to be less than 3 MHz. 

\begin{figure}
\begin{center}
\includegraphics[width=0.9\linewidth]{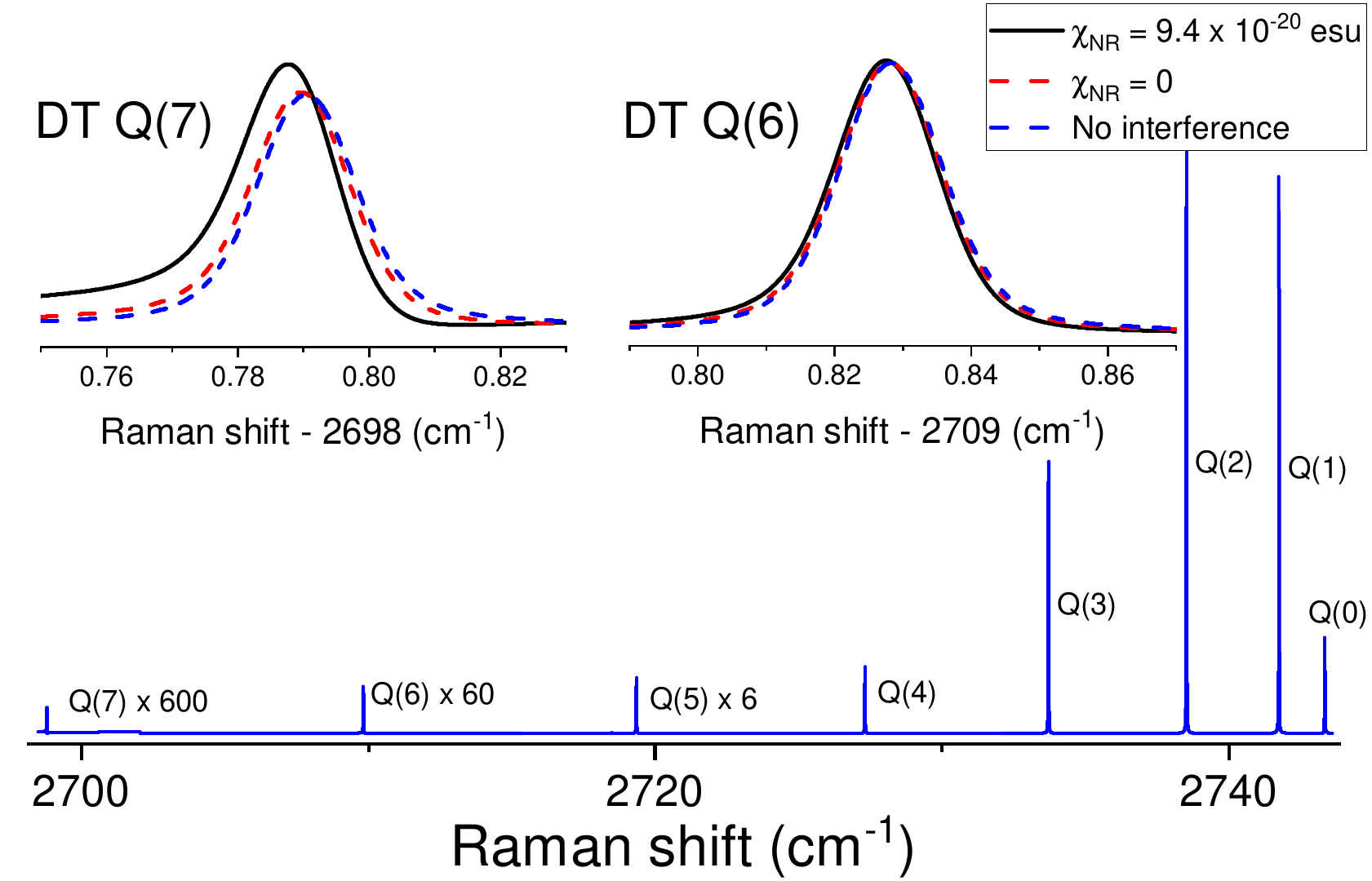}
\caption{\label{nr_simulation}
Simulation of CARS spectra of DT displaying relative intensities for Q(0-7) lines at room temperature. Zoomed-in spectra of Q(6) and Q(7) are displayed showing the resonant and non-resonant interference effects on these weak lines; a) the blue line represents a symmetric line profile without considering other contributions in the susceptibility; b) the red line represents the cross-interference from nearby line contributions in the  fundamental band Q-branch susceptibility; c) the black line represents the inclusion of both interference effects: estimated electronic nonresonant susceptibility $\chi_{NR}$ and off-resonance fundamental band Q-branch susceptibility. The shift in peak position of Q(6) and Q(7) from a) and c) is about -24 MHz and -83 MHz,  respectively.
}
\end{center}
\end{figure}

The Doppler-limited spectra for the DT lines were also simulated and depicted in Fig.~\ref{nr_simulation}.
For the strong DT transitions Q($J=0-5$), the asymmetry in the line profiles is not apparent, and the interference-induced shifts in the line positions are found to be below 3 MHz for Q($J=0-4$), while it is -8 MHz for Q(5).
However, for the weaker Q(6) and Q(7) lines, the interference effect is more visible as shown in the inset of Fig.~\ref{nr_simulation}.
Based on Eqs.~(\ref{eq:CARSchi3}) and (\ref{eq:chir}) the effects of the cross-interferences between the lines as well as the effects of the interference of the non-resonant background are quantitatively evaluated.
A shift of -24 MHz is observed for Q(6) and a much higher shift of -83 MHz for the weaker Q(7) line.
It is worth noting that for the DT Q(6) and Q(7) lines, the non-resonant susceptibility contribution $\chi_{NR}$ is an important contribution that is comparable to the cross-interference contributions, as shown in the insets of Fig.~\ref{nr_simulation}.
The $\chi_{NR}$ for the DT sample is higher than in the case of \Tm\ mainly caused by the higher total pressure in the DT gas cell.
Similar simulations were also performed on the HT Q($J=0-3$) lines, with the resulting shifts in peak positions due to CARS interference were found to be less than 3 MHz.
The line positions reported in Table~\ref{tab:Results} include corrections to these estimated CARS interference effects.

\subsection{Lineshape fitting}

A phenomenological approach is adopted in fitting asymmetric profiles to experimental data from the combined effects of the ac-Stark and spectral interference phenomena, especially for the weak transitions.
While a full quantitative assessment that would separate the contributions of these effects is not possible at present, the lineshape study enables us to better estimate the contribution to the systematic uncertainty resulting from the line profile asymmetry.

For the strong transitions with symmetric line profiles such as Q transitions from $J=0-5$ in \Tm, DT and HT, a symmetric Gaussian or Voigt fitting function is used.
However, despite the \Tm\ Q(0) and the DT Q(6) lines appearing to be symmetric, the simulations suggest that interference effects may potentially shift the line center.
To estimate a systematic uncertainty contribution due to a choice of the line profile in the fitting of the \Tm\ Q(0) and DT Q(6) lines, a Faddeeva function $w(\zeta)$  is employed:
\begin{equation} \label{eq:Faddeeva}
I_{AS}=y_0 + \left\vert\chi_{NR} - i \, a_0 \, w\left(\frac{-\omega_0+\omega_R+i\Gamma}{\Delta\omega}\right) \right\vert^{2},
\end{equation}
where $\Delta\omega$ is the Gaussian width and $\Gamma$ is the Lorentzian width, while $y_0$ sets the experimental baseline.
For \Tm\ Q(0) and DT Q(6), a 16 MHz discrepancy is observed from the fitting result between a simple symmetric Gaussian profile and the Faddeeva profile in Eq.~(\ref{eq:Faddeeva}). 
This systematic offset is included in the uncertainty budget of these two lines as listed in Table~\ref{tab:Results}.

\begin{figure}
\begin{center}
\includegraphics[width=0.9\linewidth]{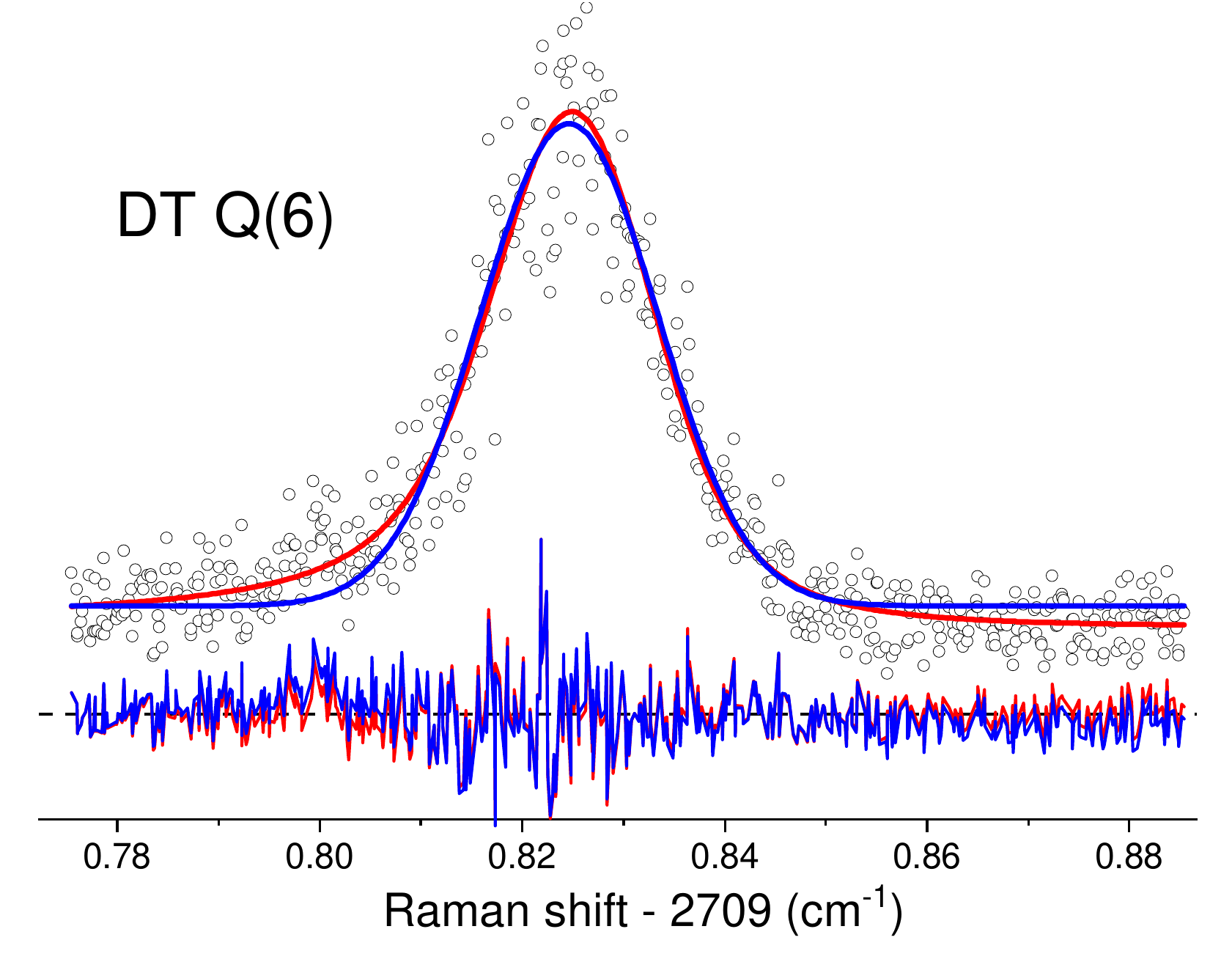}
\caption{\label{DT_Q6}
Comparison of lineshape fitting of the DT Q(6) transition using the asymmetric profile of Eq.~(\ref{eq:Faddeeva}) (red) and a symmetric Gaussian function (blue). Fitting with the Faddeeva profile, as in Eq.~(\ref{eq:Faddeeva}), gives a better fit especially in the the wing of the spectrum.
The difference of central frequency between the two fitting functions is 20 MHz.}
\end{center}
\end{figure}

\begin{figure}
\begin{center}
\includegraphics[width=0.8\linewidth]{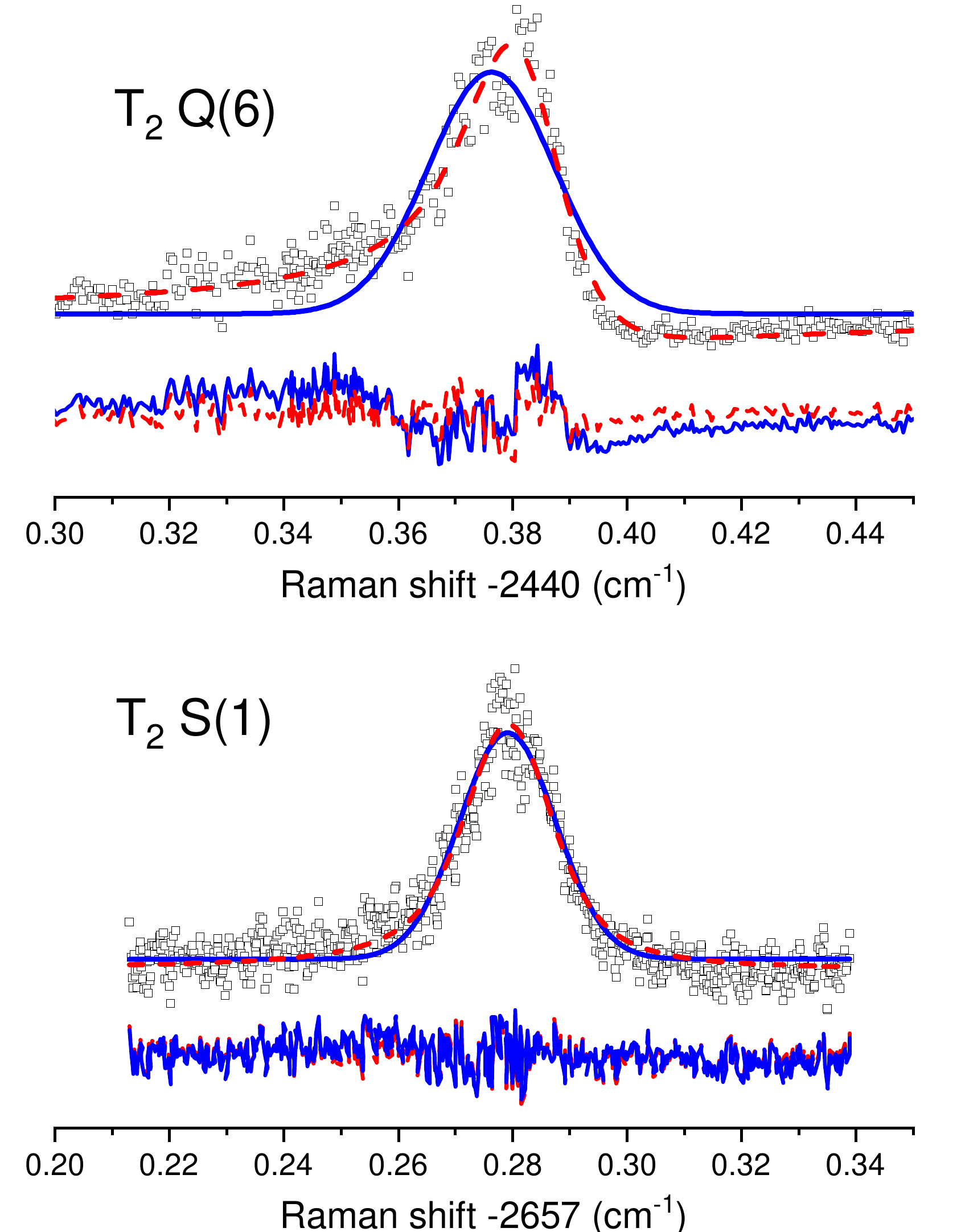}
\caption{\label{T2_Q6}
Line fitting of the \Tm\ Q(6) line and the weaker S(1) transition, both with pronounced asymmetries, using a symmetric Gaussian fit (solid - blue) and asymmetric Faddeeva profile (dashed - red).
Below the spectra the residuals of different fitting functions are plotted.}
\end{center}
\end{figure}

For other weak lines such as the \Tm\ Q(6) and S(1) transitions shown in Fig.~\ref{T2_Q6}, the poor SNR makes it more difficult to separate the effects of the dispersive feature due to spectal interference and the ac-Stark splitting and broadening, leading to larger uncertainties in the line position determination.
Note that when compared to the DT Q(6) line in Fig.~\ref{DT_Q6}, the analogous \Tm\ Q(6) line in Fig.~\ref{T2_Q6} exhibits a more pronounced asymmetry. This is due to the ortho-para spin statistics leading to a lower population for $J=6$ in homonuclear \Tm\ species, which is not present in DT. 
On the other hand, despite the \Tm\ S(1) line being of comparable strength to \Tm\ Q(6) (Fig.~\ref{T2_Q6}), the S(1) line at 2657 \wn\ exhibits a more symmetric profile. This can be explained from competing contributions of opposite sign: the cross-interference  $\chi_R$ from the \Tm\ Q-branch at 2464 \wn\ having a negative sign and the non-resonance $\chi_{NR}$ background having a positive sign. 

In addition to the cross-interference induced shifts in the peak positions (included in Table~\ref{tab:Results}), the transition frequencies of the weaker lines are corrected for the ac-Stark shift, using the slope of low-intensity ac-Stark measurements in stronger transitions (e.g. Q(1) transition of the respective species).
However, the dominant contribution to the uncertainties of $5 \times 10^{-3}$ \wn\ for the weak lines derives from the large fitting uncertainty in the peak positions given the significant ac-Stark broadening.

\section{Conclusion and outlook}
In summary, we have accurately determined transition frequencies of the fundamental vibrational band $(v=0\rightarrow 1)$ of HT, DT and \Tm. For most Q-branch transitions with $(J=0-5)$, uncertainties below $0.0005$ \wn\ are achieved, which represent more than a hundred-fold improvement over previous studies.
For the weak Q$(J=6,7)$ transitions in \Tm\ and DT and the S-branch of \Tm, the high pulse intensities required for the measurements cause significant ac-Stark broadening. The transition frequencies of these weak lines are thus reported with a much higher uncertainty of $0.005$ \wn.
All measured transition energies are in very good agreement with the latest \emph{ab initio} calculations based on NAPT theoretical framework~\cite{Czachorowski2018,SPECTRE}. This work demonstrates that measurements on tritiated species can be achieved at high accuracy, to within a factor of three compared to the accuracy for the non-radioactive species. This doubles the number of benchmark candidates for testing the most accurate quantum chemical calculations of the energy structure of molecular hydrogen. 

Due to interference effects with the non-resonant background in CARS spectra, the resulting asymmetric line profiles in weak transitions lead to a higher uncertainty in the line center determination.
Cross-interference between resonances can be a limiting factor for the spectroscopic accuracy of CARS measurements; it is shown here that this phenomenon can be quantitatively approached and corrections to transition frequencies calculated.

Future improvements in experiments on tritium-bearing molecular hydrogen would require sub-Doppler techniques, to enable improved comparison with ever advancing calculations.
This could be achieved by exploiting Lamb dip spectroscopy with hybrid pulsed-cw CARS, using a pulsed pump beam and a continuous wave Stokes beam, as demonstrated here for \Dm. From these test studies it follows that an improved understanding of line shape profiles, in particular of the ac-Stark effect on Lamb-dips, is required to effectively achieve higher accuracy on the transition frequencies. 
Improvements might be achieved by using higher cw-Stokes powers, allowing measurements to be performed at even lower pump pulse energies, thereby reducing ac-Stark effects induced by the pump laser.
With the narrower cw laser bandwidth and improved frequency calibration procedures, sub-MHz accuracies could potentially be achieved.
Ultimately the application of novel cavity-enhanced cw-spectroscopies such as the NICE-OHMS technique on the heteronuclear species HT and DT, with the possibility of Doppler-free saturated absorption, may result in accuracy improvements by orders of magnitude~\cite{Cozijn2018}, provided that those techniques can be combined with safe handling of radioactive species.

\section*{Conflicts of interest}
There are no conflicts to declare.

\section*{Acknowledgements}
Pawe\l{} Czachorowski, Krzysztof Pachucki, Jacek Komasa and Mariusz Puchalski are acknowledged for sharing results on theoretical NAPT calculations. We are grateful to Tobias Falke, David Hillesheimer, Stefan Welte and J\"urgen Wendel of TLK for the preparation and handling of the tritium cell and gas samples. WU thanks the European Research Council for an ERC-Advanced grant (No 670168).
The research leading to these results has received funding from LASERLAB-EUROPE (grant agreement no. 654148, European Union’s Horizon 2020 research and innovation programme).



\balance


\bibliography{Hydrogen,Tritium,CARS,Spectroscopy-general} 

\providecommand*{\mcitethebibliography}{\thebibliography}
\csname @ifundefined\endcsname{endmcitethebibliography}
{\let\endmcitethebibliography\endthebibliography}{}
\begin{mcitethebibliography}{80}
\providecommand*{\natexlab}[1]{#1}
\providecommand*{\mciteSetBstSublistMode}[1]{}
\providecommand*{\mciteSetBstMaxWidthForm}[2]{}
\providecommand*{\mciteBstWouldAddEndPuncttrue}
  {\def\EndOfBibitem{\unskip.}}
\providecommand*{\mciteBstWouldAddEndPunctfalse}
  {\let\EndOfBibitem\relax}
\providecommand*{\mciteSetBstMidEndSepPunct}[3]{}
\providecommand*{\mciteSetBstSublistLabelBeginEnd}[3]{}
\providecommand*{\EndOfBibitem}{}
\mciteSetBstSublistMode{f}
\mciteSetBstMaxWidthForm{subitem}
{(\emph{\alph{mcitesubitemcount}})}
\mciteSetBstSublistLabelBeginEnd{\mcitemaxwidthsubitemform\space}
{\relax}{\relax}

\bibitem[Salumbides \emph{et~al.}(2013)Salumbides, Koelemeij, Komasa, Pachucki,
  Eikema, and Ubachs]{Salumbides2013}
E.~J. Salumbides, J.~C.~J. Koelemeij, J.~Komasa, K.~Pachucki, K.~S.~E. Eikema
  and W.~Ubachs, \emph{Phys. Rev. D}, 2013, \textbf{87}, 112008\relax
\mciteBstWouldAddEndPuncttrue
\mciteSetBstMidEndSepPunct{\mcitedefaultmidpunct}
{\mcitedefaultendpunct}{\mcitedefaultseppunct}\relax
\EndOfBibitem
\bibitem[Salumbides \emph{et~al.}(2015)Salumbides, Schellekens, Gato-Rivera,
  and Ubachs]{Salumbides2015b}
E.~J. Salumbides, A.~N. Schellekens, B.~Gato-Rivera and W.~Ubachs, \emph{New J.
  Phys.}, 2015, \textbf{17}, 033015\relax
\mciteBstWouldAddEndPuncttrue
\mciteSetBstMidEndSepPunct{\mcitedefaultmidpunct}
{\mcitedefaultendpunct}{\mcitedefaultseppunct}\relax
\EndOfBibitem
\bibitem[Ubachs \emph{et~al.}(2016)Ubachs, Koelemeij, Eikema, and
  Salumbides]{Ubachs2016}
W.~Ubachs, J.~C.~J. Koelemeij, K.~S.~E. Eikema and E.~J. Salumbides, \emph{J.
  Mol. Spectrosc.}, 2016, \textbf{320}, 1 -- 12\relax
\mciteBstWouldAddEndPuncttrue
\mciteSetBstMidEndSepPunct{\mcitedefaultmidpunct}
{\mcitedefaultendpunct}{\mcitedefaultseppunct}\relax
\EndOfBibitem
\bibitem[Liu \emph{et~al.}(2009)Liu, Salumbides, Hollenstein, Koelemeij,
  Eikema, Ubachs, and Merkt]{Liu2009}
J.~Liu, E.~J. Salumbides, U.~Hollenstein, J.~C.~J. Koelemeij, K.~S.~E. Eikema,
  W.~Ubachs and F.~Merkt, \emph{J. Chem. Phys.}, 2009, \textbf{130},
  174306\relax
\mciteBstWouldAddEndPuncttrue
\mciteSetBstMidEndSepPunct{\mcitedefaultmidpunct}
{\mcitedefaultendpunct}{\mcitedefaultseppunct}\relax
\EndOfBibitem
\bibitem[Cheng \emph{et~al.}(2018)Cheng, Hussels, Niu, Bethlem, Eikema,
  Salumbides, Ubachs, Beyer, H\"{o}lsch, Agner, Merkt, Tao, Hu, and
  Jungen]{Cheng2018}
C.-F. Cheng, J.~Hussels, M.~Niu, H.~L. Bethlem, K.~S.~E. Eikema, E.~J.
  Salumbides, W.~Ubachs, M.~Beyer, N.~H\"{o}lsch, J.~A. Agner, F.~Merkt, L.-G.
  Tao, S.-M. Hu and C.~Jungen, \emph{Phys. Rev. Lett.}, 2018, \textbf{121},
  013001\relax
\mciteBstWouldAddEndPuncttrue
\mciteSetBstMidEndSepPunct{\mcitedefaultmidpunct}
{\mcitedefaultendpunct}{\mcitedefaultseppunct}\relax
\EndOfBibitem
\bibitem[H\"{o}lsch \emph{et~al.}(2019)H\"{o}lsch, Beyer, Salumbides, Eikema,
  Ubachs, Jungen, and Merkt]{Holsch2019}
N.~H\"{o}lsch, M.~Beyer, E.~J. Salumbides, K.~S.~E. Eikema, W.~Ubachs,
  C.~Jungen and F.~Merkt, \emph{Phys. Rev. Lett.}, 2019, \textbf{122},
  103002\relax
\mciteBstWouldAddEndPuncttrue
\mciteSetBstMidEndSepPunct{\mcitedefaultmidpunct}
{\mcitedefaultendpunct}{\mcitedefaultseppunct}\relax
\EndOfBibitem
\bibitem[Beyer \emph{et~al.}(2019)Beyer, H\"olsch, Hussels, Cheng, Salumbides,
  Eikema, Ubachs, Jungen, and Merkt]{Beyer2019}
M.~Beyer, N.~H\"olsch, J.~Hussels, C.-F. Cheng, E.~J. Salumbides, K.~S.~E.
  Eikema, W.~Ubachs, C.~Jungen and F.~Merkt, \emph{Phys. Rev. Lett.}, 2019,
  \textbf{123}, 163002\relax
\mciteBstWouldAddEndPuncttrue
\mciteSetBstMidEndSepPunct{\mcitedefaultmidpunct}
{\mcitedefaultendpunct}{\mcitedefaultseppunct}\relax
\EndOfBibitem
\bibitem[Dickenson \emph{et~al.}(2013)Dickenson, Niu, Salumbides, Komasa,
  Eikema, Pachucki, and Ubachs]{Dickenson2013}
G.~D. Dickenson, M.~L. Niu, E.~J. Salumbides, J.~Komasa, K.~S.~E. Eikema,
  K.~Pachucki and W.~Ubachs, \emph{Phys. Rev. lett.}, 2013, \textbf{110},
  193601\relax
\mciteBstWouldAddEndPuncttrue
\mciteSetBstMidEndSepPunct{\mcitedefaultmidpunct}
{\mcitedefaultendpunct}{\mcitedefaultseppunct}\relax
\EndOfBibitem
\bibitem[Campargue \emph{et~al.}(2012)Campargue, Kassi, Pachucki, and
  Komasa]{Campargue2012}
A.~Campargue, S.~Kassi, K.~Pachucki and J.~Komasa, \emph{Phys. Chem. Chem.
  Phys.}, 2012, \textbf{14}, 802--815\relax
\mciteBstWouldAddEndPuncttrue
\mciteSetBstMidEndSepPunct{\mcitedefaultmidpunct}
{\mcitedefaultendpunct}{\mcitedefaultseppunct}\relax
\EndOfBibitem
\bibitem[Kassi and Campargue(2014)]{Kassi2014}
S.~Kassi and A.~Campargue, \emph{J. Mol. Spectr.}, 2014, \textbf{300}, 55 --
  59\relax
\mciteBstWouldAddEndPuncttrue
\mciteSetBstMidEndSepPunct{\mcitedefaultmidpunct}
{\mcitedefaultendpunct}{\mcitedefaultseppunct}\relax
\EndOfBibitem
\bibitem[Cheng \emph{et~al.}(2012)Cheng, Sun, Pan, Wang, Liu, Campargue, and
  Hu]{Cheng2012}
C.-F. Cheng, Y.~R. Sun, H.~Pan, J.~Wang, A.-W. Liu, A.~Campargue and S.-M. Hu,
  \emph{Phys. Rev. A}, 2012, \textbf{85}, 024501\relax
\mciteBstWouldAddEndPuncttrue
\mciteSetBstMidEndSepPunct{\mcitedefaultmidpunct}
{\mcitedefaultendpunct}{\mcitedefaultseppunct}\relax
\EndOfBibitem
\bibitem[Maddaloni \emph{et~al.}(2010)Maddaloni, Malara, {De Tommasi}, {De
  Rosa}, Ricciardi, Gagliardi, Tamassia, {Di Lonardo}, and {De
  Natale}]{Maddaloni2010}
P.~Maddaloni, P.~Malara, E.~{De Tommasi}, M.~{De Rosa}, I.~Ricciardi,
  G.~Gagliardi, F.~Tamassia, G.~{Di Lonardo} and P.~{De Natale}, \emph{J. Chem.
  Phys.}, 2010, \textbf{133}, 154317\relax
\mciteBstWouldAddEndPuncttrue
\mciteSetBstMidEndSepPunct{\mcitedefaultmidpunct}
{\mcitedefaultendpunct}{\mcitedefaultseppunct}\relax
\EndOfBibitem
\bibitem[Kassi \emph{et~al.}(2012)Kassi, Campargue, Pachucki, and
  Komasa]{Kassi2012}
S.~Kassi, A.~Campargue, K.~Pachucki and J.~Komasa, \emph{J. Chem. Phys.}, 2012,
  \textbf{136}, 184309\relax
\mciteBstWouldAddEndPuncttrue
\mciteSetBstMidEndSepPunct{\mcitedefaultmidpunct}
{\mcitedefaultendpunct}{\mcitedefaultseppunct}\relax
\EndOfBibitem
\bibitem[Kassi and Campargue(2011)]{Kassi2011}
S.~Kassi and A.~Campargue, \emph{J. Mol. Spectrosc.}, 2011, \textbf{267},
  36--42\relax
\mciteBstWouldAddEndPuncttrue
\mciteSetBstMidEndSepPunct{\mcitedefaultmidpunct}
{\mcitedefaultendpunct}{\mcitedefaultseppunct}\relax
\EndOfBibitem
\bibitem[Tao \emph{et~al.}(2018)Tao, Liu, Pachucki, Komasa, Sun, Wang, and
  Hu]{Tao2018}
L.-G. Tao, A.-W. Liu, K.~Pachucki, J.~Komasa, Y.~R. Sun, J.~Wang and S.-M. Hu,
  \emph{Phys. Rev. Lett.}, 2018, \textbf{120}, 153001\relax
\mciteBstWouldAddEndPuncttrue
\mciteSetBstMidEndSepPunct{\mcitedefaultmidpunct}
{\mcitedefaultendpunct}{\mcitedefaultseppunct}\relax
\EndOfBibitem
\bibitem[Cozijn \emph{et~al.}(2018)Cozijn, Dupr\'{e}, Salumbides, Eikema, and
  Ubachs]{Cozijn2018}
F.~M.~J. Cozijn, P.~Dupr\'{e}, E.~J. Salumbides, K.~S.~E. Eikema and W.~Ubachs,
  \emph{Phys. Rev. Lett.}, 2018, \textbf{120}, 153002\relax
\mciteBstWouldAddEndPuncttrue
\mciteSetBstMidEndSepPunct{\mcitedefaultmidpunct}
{\mcitedefaultendpunct}{\mcitedefaultseppunct}\relax
\EndOfBibitem
\bibitem[Fasci \emph{et~al.}(2018)Fasci, Castrillo, Dinesan, Gravina, Moretti,
  and Gianfrani]{Fasci2018}
E.~Fasci, A.~Castrillo, H.~Dinesan, S.~Gravina, L.~Moretti and L.~Gianfrani,
  \emph{Phys. Rev. A}, 2018, \textbf{98}, 022516\relax
\mciteBstWouldAddEndPuncttrue
\mciteSetBstMidEndSepPunct{\mcitedefaultmidpunct}
{\mcitedefaultendpunct}{\mcitedefaultseppunct}\relax
\EndOfBibitem
\bibitem[Komasa \emph{et~al.}(2011)Komasa, Piszczatowski, \L{}ach, Przybytek,
  Jeziorski, and Pachucki]{Komasa2011}
J.~Komasa, K.~Piszczatowski, G.~\L{}ach, M.~Przybytek, B.~Jeziorski and
  K.~Pachucki, \emph{J. Chem. Theory Comput.}, 2011, \textbf{7},
  3105--3115\relax
\mciteBstWouldAddEndPuncttrue
\mciteSetBstMidEndSepPunct{\mcitedefaultmidpunct}
{\mcitedefaultendpunct}{\mcitedefaultseppunct}\relax
\EndOfBibitem
\bibitem[Pachucki and Komasa(2010)]{Pachucki2010b}
K.~Pachucki and J.~Komasa, \emph{Phys. Chem. Chem. Phys.}, 2010, \textbf{12},
  9188--9196\relax
\mciteBstWouldAddEndPuncttrue
\mciteSetBstMidEndSepPunct{\mcitedefaultmidpunct}
{\mcitedefaultendpunct}{\mcitedefaultseppunct}\relax
\EndOfBibitem
\bibitem[Pachucki(2010)]{Pachucki2010}
K.~Pachucki, \emph{Phys. Rev. A}, 2010, \textbf{82}, 032509\relax
\mciteBstWouldAddEndPuncttrue
\mciteSetBstMidEndSepPunct{\mcitedefaultmidpunct}
{\mcitedefaultendpunct}{\mcitedefaultseppunct}\relax
\EndOfBibitem
\bibitem[Pachucki and Komasa(2014)]{Pachucki2014}
K.~Pachucki and J.~Komasa, \emph{J. Chem. Phys.}, 2014, \textbf{141},
  224103\relax
\mciteBstWouldAddEndPuncttrue
\mciteSetBstMidEndSepPunct{\mcitedefaultmidpunct}
{\mcitedefaultendpunct}{\mcitedefaultseppunct}\relax
\EndOfBibitem
\bibitem[Pachucki and Komasa(2015)]{Pachucki2015}
K.~Pachucki and J.~Komasa, \emph{J. Chem. Phys.}, 2015, \textbf{143},
  034111\relax
\mciteBstWouldAddEndPuncttrue
\mciteSetBstMidEndSepPunct{\mcitedefaultmidpunct}
{\mcitedefaultendpunct}{\mcitedefaultseppunct}\relax
\EndOfBibitem
\bibitem[Puchalski \emph{et~al.}(2017)Puchalski, Komasa, and
  Pachucki]{Puchalski2017}
M.~Puchalski, J.~Komasa and K.~Pachucki, \emph{Phys. Rev. A}, 2017,
  \textbf{95}, 052506\relax
\mciteBstWouldAddEndPuncttrue
\mciteSetBstMidEndSepPunct{\mcitedefaultmidpunct}
{\mcitedefaultendpunct}{\mcitedefaultseppunct}\relax
\EndOfBibitem
\bibitem[Puchalski \emph{et~al.}(2016)Puchalski, Komasa, Czachorowski, and
  Pachucki]{Puchalski2016}
M.~Puchalski, J.~Komasa, P.~Czachorowski and K.~Pachucki, \emph{Phys. Rev.
  Lett.}, 2016, \textbf{117}, 263002\relax
\mciteBstWouldAddEndPuncttrue
\mciteSetBstMidEndSepPunct{\mcitedefaultmidpunct}
{\mcitedefaultendpunct}{\mcitedefaultseppunct}\relax
\EndOfBibitem
\bibitem[Czachorowski \emph{et~al.}(2018)Czachorowski, Puchalski, Komasa, and
  Pachucki]{Czachorowski2018}
P.~Czachorowski, M.~Puchalski, J.~Komasa and K.~Pachucki, \emph{Phys. Rev. A},
  2018, \textbf{98}, 052506\relax
\mciteBstWouldAddEndPuncttrue
\mciteSetBstMidEndSepPunct{\mcitedefaultmidpunct}
{\mcitedefaultendpunct}{\mcitedefaultseppunct}\relax
\EndOfBibitem
\bibitem[Komasa \emph{et~al.}(2019)Komasa, Puchalski, Czachorowski, \L{}ach,
  and Pachucki]{Komasa2019}
J.~Komasa, M.~Puchalski, P.~Czachorowski, G.~\L{}ach and K.~Pachucki,
  \emph{Phys. Rev. A}, 2019, \textbf{100}, 032519\relax
\mciteBstWouldAddEndPuncttrue
\mciteSetBstMidEndSepPunct{\mcitedefaultmidpunct}
{\mcitedefaultendpunct}{\mcitedefaultseppunct}\relax
\EndOfBibitem
\bibitem[Pachucki and Komasa(2018)]{Pachucki2018}
K.~Pachucki and J.~Komasa, \emph{Phys. Chem. Chem. Phys.}, 2018, \textbf{20},
  247--255\relax
\mciteBstWouldAddEndPuncttrue
\mciteSetBstMidEndSepPunct{\mcitedefaultmidpunct}
{\mcitedefaultendpunct}{\mcitedefaultseppunct}\relax
\EndOfBibitem
\bibitem[Puchalski \emph{et~al.}(2019)Puchalski, Komasa, Spyszkiewicz, and
  Pachucki]{Puchalski2019}
M.~Puchalski, J.~Komasa, A.~Spyszkiewicz and K.~Pachucki, \emph{Phys. Rev. A},
  2019, \textbf{100}, 020503\relax
\mciteBstWouldAddEndPuncttrue
\mciteSetBstMidEndSepPunct{\mcitedefaultmidpunct}
{\mcitedefaultendpunct}{\mcitedefaultseppunct}\relax
\EndOfBibitem
\bibitem[Dieke and Tomkins(1949)]{Dieke1949}
G.~H. Dieke and F.~S. Tomkins, \emph{Phys. Rev.}, 1949, \textbf{76},
  283--289\relax
\mciteBstWouldAddEndPuncttrue
\mciteSetBstMidEndSepPunct{\mcitedefaultmidpunct}
{\mcitedefaultendpunct}{\mcitedefaultseppunct}\relax
\EndOfBibitem
\bibitem[Dieke(1958)]{Dieke1958}
G.~Dieke, \emph{J. Mol. Spectr.}, 1958, \textbf{2}, 494 -- 517\relax
\mciteBstWouldAddEndPuncttrue
\mciteSetBstMidEndSepPunct{\mcitedefaultmidpunct}
{\mcitedefaultendpunct}{\mcitedefaultseppunct}\relax
\EndOfBibitem
\bibitem[Edwards \emph{et~al.}(1978)Edwards, Long, and Mansour]{Edwards1978}
H.~G.~M. Edwards, D.~A. Long and H.~R. Mansour, \emph{J. Chem. Soc.{,} Faraday
  Trans. 2}, 1978, \textbf{74}, 1203--1207\relax
\mciteBstWouldAddEndPuncttrue
\mciteSetBstMidEndSepPunct{\mcitedefaultmidpunct}
{\mcitedefaultendpunct}{\mcitedefaultseppunct}\relax
\EndOfBibitem
\bibitem[Edwards \emph{et~al.}(1979)Edwards, Long, Mansour, and
  Najm]{Edwards1979}
H.~G.~M. Edwards, D.~A. Long, H.~R. Mansour and K.~A.~B. Najm, \emph{J. Raman
  Spectrosc.}, 1979, \textbf{8}, 251--254\relax
\mciteBstWouldAddEndPuncttrue
\mciteSetBstMidEndSepPunct{\mcitedefaultmidpunct}
{\mcitedefaultendpunct}{\mcitedefaultseppunct}\relax
\EndOfBibitem
\bibitem[{Veirs} and {Rosenblatt}(1987)]{Veirs1987}
D.~K. {Veirs} and G.~M. {Rosenblatt}, \emph{J. Mol. Spectrosc.}, 1987,
  \textbf{121}, 401--419\relax
\mciteBstWouldAddEndPuncttrue
\mciteSetBstMidEndSepPunct{\mcitedefaultmidpunct}
{\mcitedefaultendpunct}{\mcitedefaultseppunct}\relax
\EndOfBibitem
\bibitem[Chuang and Zare(1987)]{Chuang1987}
M.-C. Chuang and R.~N. Zare, \emph{J. Mol. Spectrosc.}, 1987, \textbf{121}, 380
  -- 400\relax
\mciteBstWouldAddEndPuncttrue
\mciteSetBstMidEndSepPunct{\mcitedefaultmidpunct}
{\mcitedefaultendpunct}{\mcitedefaultseppunct}\relax
\EndOfBibitem
\bibitem[Sprecher \emph{et~al.}(2010)Sprecher, Liu, Jungen, Ubachs, and
  Merkt]{Sprecher2010}
D.~Sprecher, J.~Liu, C.~Jungen, W.~Ubachs and F.~Merkt, \emph{J. Chem. Phys.},
  2010, \textbf{133}, 111102\relax
\mciteBstWouldAddEndPuncttrue
\mciteSetBstMidEndSepPunct{\mcitedefaultmidpunct}
{\mcitedefaultendpunct}{\mcitedefaultseppunct}\relax
\EndOfBibitem
\bibitem[Aker \emph{et~al.}(2019)Aker\emph{et~al.}]{Aker2019}
M.~Aker \emph{et~al.}, \emph{Phys. Rev. Lett.}, 2019, \textbf{123},
  221802\relax
\mciteBstWouldAddEndPuncttrue
\mciteSetBstMidEndSepPunct{\mcitedefaultmidpunct}
{\mcitedefaultendpunct}{\mcitedefaultseppunct}\relax
\EndOfBibitem
\bibitem[Saenz \emph{et~al.}(2000)Saenz, Jonsell, and Froelich]{Saenz2000}
A.~Saenz, S.~Jonsell and P.~Froelich, \emph{Phys. Rev. Lett.}, 2000,
  \textbf{84}, 242--245\relax
\mciteBstWouldAddEndPuncttrue
\mciteSetBstMidEndSepPunct{\mcitedefaultmidpunct}
{\mcitedefaultendpunct}{\mcitedefaultseppunct}\relax
\EndOfBibitem
\bibitem[Doss \emph{et~al.}(2006)Doss, Tennyson, Saenz, and Jonsell]{Doss2006}
N.~Doss, J.~Tennyson, A.~Saenz and S.~Jonsell, \emph{Phys. Rev. C}, 2006,
  \textbf{73}, 025502\relax
\mciteBstWouldAddEndPuncttrue
\mciteSetBstMidEndSepPunct{\mcitedefaultmidpunct}
{\mcitedefaultendpunct}{\mcitedefaultseppunct}\relax
\EndOfBibitem
\bibitem[Bodine \emph{et~al.}(2015)Bodine, Parno, and Robertson]{Bodine2015}
L.~I. Bodine, D.~Parno and R.~Robertson, \emph{Phys. Rev. C}, 2015,
  \textbf{91}, 035505\relax
\mciteBstWouldAddEndPuncttrue
\mciteSetBstMidEndSepPunct{\mcitedefaultmidpunct}
{\mcitedefaultendpunct}{\mcitedefaultseppunct}\relax
\EndOfBibitem
\bibitem[Kleesiek \emph{et~al.}(2019)Kleesiek, Behrens, Drexlin, Eitel, Erhard,
  Formaggio, Glück, Groh, Hötzel, Mertens, Poon, Weinheimer, and
  Valerius]{Kleesiek2019}
M.~Kleesiek, J.~Behrens, G.~Drexlin, K.~Eitel, M.~Erhard, J.~A. Formaggio,
  F.~Glück, S.~Groh, M.~Hötzel, S.~Mertens, A.~W.~P. Poon, C.~Weinheimer and
  K.~Valerius, \emph{Eur. Phys. J. C}, 2019, \textbf{79}, 204\relax
\mciteBstWouldAddEndPuncttrue
\mciteSetBstMidEndSepPunct{\mcitedefaultmidpunct}
{\mcitedefaultendpunct}{\mcitedefaultseppunct}\relax
\EndOfBibitem
\bibitem[James \emph{et~al.}(2013)James, Schl\"{o}sser, Fischer, Sturm,
  Bornschein, Lewis, and Telle]{James2013}
T.~M. James, M.~Schl\"{o}sser, S.~Fischer, M.~Sturm, B.~Bornschein, R.~J. Lewis
  and H.~H. Telle, \emph{J. Raman Spectr.}, 2013, \textbf{44}, 857--865\relax
\mciteBstWouldAddEndPuncttrue
\mciteSetBstMidEndSepPunct{\mcitedefaultmidpunct}
{\mcitedefaultendpunct}{\mcitedefaultseppunct}\relax
\EndOfBibitem
\bibitem[Schlösser \emph{et~al.}(2013)Schlösser, Seitz, Rupp, Herwig, Alecu,
  Sturm, and Bornschein]{Schlosser2013}
M.~Schlösser, H.~Seitz, S.~Rupp, P.~Herwig, C.~G. Alecu, M.~Sturm and
  B.~Bornschein, \emph{Anal. Chem.}, 2013, \textbf{85}, 2739--2745\relax
\mciteBstWouldAddEndPuncttrue
\mciteSetBstMidEndSepPunct{\mcitedefaultmidpunct}
{\mcitedefaultendpunct}{\mcitedefaultseppunct}\relax
\EndOfBibitem
\bibitem[Schl\"{o}sser \emph{et~al.}(2015)Schl\"{o}sser, Bornschein, Fischer,
  Kassel, Rupp, Sturm, James, and Telle]{Schloesser2015}
M.~Schl\"{o}sser, B.~Bornschein, S.~Fischer, F.~Kassel, S.~Rupp, M.~Sturm,
  T.~James and H.~Telle, \emph{Fusion Science and Technology}, 2015,
  \textbf{67}, 555--558\relax
\mciteBstWouldAddEndPuncttrue
\mciteSetBstMidEndSepPunct{\mcitedefaultmidpunct}
{\mcitedefaultendpunct}{\mcitedefaultseppunct}\relax
\EndOfBibitem
\bibitem[Schl\"{o}sser \emph{et~al.}(2017)Schl\"{o}sser, Zhao, Trivikram,
  Ubachs, and Salumbides]{Schlosser2017}
M.~Schl\"{o}sser, X.~Zhao, M.~T. Trivikram, W.~Ubachs and E.~J. Salumbides,
  \emph{J. Phys. B}, 2017, \textbf{50}, 214004\relax
\mciteBstWouldAddEndPuncttrue
\mciteSetBstMidEndSepPunct{\mcitedefaultmidpunct}
{\mcitedefaultendpunct}{\mcitedefaultseppunct}\relax
\EndOfBibitem
\bibitem[Trivikram \emph{et~al.}(2018)Trivikram, Schl\"{o}sser, Ubachs, and
  Salumbides]{Trivikram2018}
T.~M. Trivikram, M.~Schl\"{o}sser, W.~Ubachs and E.~J. Salumbides, \emph{Phys.
  Rev. Lett.}, 2018, \textbf{120}, 163002\relax
\mciteBstWouldAddEndPuncttrue
\mciteSetBstMidEndSepPunct{\mcitedefaultmidpunct}
{\mcitedefaultendpunct}{\mcitedefaultseppunct}\relax
\EndOfBibitem
\bibitem[Lai \emph{et~al.}(2019)Lai, Czachorowski, Schl\"osser, Puchalski,
  Komasa, Pachucki, Ubachs, and Salumbides]{Lai2019}
K.-F. Lai, P.~Czachorowski, M.~Schl\"osser, M.~Puchalski, J.~Komasa,
  K.~Pachucki, W.~Ubachs and E.~J. Salumbides, \emph{Phys. Rev. Research},
  2019, \textbf{1}, 033124\relax
\mciteBstWouldAddEndPuncttrue
\mciteSetBstMidEndSepPunct{\mcitedefaultmidpunct}
{\mcitedefaultendpunct}{\mcitedefaultseppunct}\relax
\EndOfBibitem
\bibitem[Altmann \emph{et~al.}(2018)Altmann, Dreissen, Salumbides, Ubachs, and
  Eikema]{Altmann2018}
R.~K. Altmann, L.~S. Dreissen, E.~J. Salumbides, W.~Ubachs and K.~S.~E. Eikema,
  \emph{Phys. Rev. Lett.}, 2018, \textbf{120}, 043204\relax
\mciteBstWouldAddEndPuncttrue
\mciteSetBstMidEndSepPunct{\mcitedefaultmidpunct}
{\mcitedefaultendpunct}{\mcitedefaultseppunct}\relax
\EndOfBibitem
\bibitem[Sazonov and Magomedbekov(2011)]{Sazonov2011}
A.~B. Sazonov and E.~P. Magomedbekov, \emph{Fusion Sci. Technol.}, 2011,
  \textbf{60}, 1383--1386\relax
\mciteBstWouldAddEndPuncttrue
\mciteSetBstMidEndSepPunct{\mcitedefaultmidpunct}
{\mcitedefaultendpunct}{\mcitedefaultseppunct}\relax
\EndOfBibitem
\bibitem[Jones(1967)]{Jones1967}
W.~M. Jones, \emph{J. Chem. Phys.}, 1967, \textbf{47}, 4675--4679\relax
\mciteBstWouldAddEndPuncttrue
\mciteSetBstMidEndSepPunct{\mcitedefaultmidpunct}
{\mcitedefaultendpunct}{\mcitedefaultseppunct}\relax
\EndOfBibitem
\bibitem[Tolles \emph{et~al.}(1977)Tolles, Nibler, McDonald, and
  Harvey]{Tolles1977}
W.~M. Tolles, J.~W. Nibler, J.~R. McDonald and A.~B. Harvey, \emph{Appl.
  Spectrosc.}, 1977, \textbf{31}, 253--271\relax
\mciteBstWouldAddEndPuncttrue
\mciteSetBstMidEndSepPunct{\mcitedefaultmidpunct}
{\mcitedefaultendpunct}{\mcitedefaultseppunct}\relax
\EndOfBibitem
\bibitem[Eikema \emph{et~al.}(1997)Eikema, Ubachs, Vassen, and
  Hogervorst]{Eikema1997}
K.~S.~E. Eikema, W.~Ubachs, W.~Vassen and W.~Hogervorst, \emph{Phys. Rev. A},
  1997, \textbf{55}, 1866\relax
\mciteBstWouldAddEndPuncttrue
\mciteSetBstMidEndSepPunct{\mcitedefaultmidpunct}
{\mcitedefaultendpunct}{\mcitedefaultseppunct}\relax
\EndOfBibitem
\bibitem[Lucht and Farrow(1988)]{Lucht1988}
R.~P. Lucht and R.~L. Farrow, \emph{J. Opt. Soc. Am. B}, 1988, \textbf{5},
  1243--1252\relax
\mciteBstWouldAddEndPuncttrue
\mciteSetBstMidEndSepPunct{\mcitedefaultmidpunct}
{\mcitedefaultendpunct}{\mcitedefaultseppunct}\relax
\EndOfBibitem
\bibitem[Xu \emph{et~al.}(2000)Xu, van Dierendonck, Hogervorst, and
  Ubachs]{Xu2000}
S.~Xu, R.~van Dierendonck, W.~Hogervorst and W.~Ubachs, \emph{J. Mol. Spectr.},
  2000, \textbf{201}, 256--266\relax
\mciteBstWouldAddEndPuncttrue
\mciteSetBstMidEndSepPunct{\mcitedefaultmidpunct}
{\mcitedefaultendpunct}{\mcitedefaultseppunct}\relax
\EndOfBibitem
\bibitem[Fee \emph{et~al.}(1992)Fee, Danzmann, and Chu]{Fee1992}
M.~S. Fee, K.~Danzmann and S.~Chu, \emph{Phys. Rev. A}, 1992, \textbf{45},
  4911--4924\relax
\mciteBstWouldAddEndPuncttrue
\mciteSetBstMidEndSepPunct{\mcitedefaultmidpunct}
{\mcitedefaultendpunct}{\mcitedefaultseppunct}\relax
\EndOfBibitem
\bibitem[Gangopadhyay \emph{et~al.}(1994)Gangopadhyay, Melikechi, and
  Eyler]{Gangopadhyay1994}
S.~Gangopadhyay, N.~Melikechi and E.~E. Eyler, \emph{J. Opt. Soc. Am. B}, 1994,
  \textbf{11}, 231--241\relax
\mciteBstWouldAddEndPuncttrue
\mciteSetBstMidEndSepPunct{\mcitedefaultmidpunct}
{\mcitedefaultendpunct}{\mcitedefaultseppunct}\relax
\EndOfBibitem
\bibitem[Hannemann \emph{et~al.}(2007)Hannemann, van Duijn, and
  Ubachs]{Hannemann2007a}
S.~Hannemann, E.-J. van Duijn and W.~Ubachs, \emph{Rev. Scient. Instrum.},
  2007, \textbf{78}, 103102\relax
\mciteBstWouldAddEndPuncttrue
\mciteSetBstMidEndSepPunct{\mcitedefaultmidpunct}
{\mcitedefaultendpunct}{\mcitedefaultseppunct}\relax
\EndOfBibitem
\bibitem[Stellmer and Schreck(2014)]{Stellmer2014}
S.~Stellmer and F.~Schreck, \emph{Phys. Rev. A}, 2014, \textbf{90},
  022512\relax
\mciteBstWouldAddEndPuncttrue
\mciteSetBstMidEndSepPunct{\mcitedefaultmidpunct}
{\mcitedefaultendpunct}{\mcitedefaultseppunct}\relax
\EndOfBibitem
\bibitem[Rahn \emph{et~al.}(1980)Rahn, Farrow, Koszykowski, and
  Mattern]{Rahn1980}
L.~A. Rahn, R.~L. Farrow, M.~L. Koszykowski and P.~L. Mattern, \emph{Phys. Rev.
  Lett.}, 1980, \textbf{45}, 620--623\relax
\mciteBstWouldAddEndPuncttrue
\mciteSetBstMidEndSepPunct{\mcitedefaultmidpunct}
{\mcitedefaultendpunct}{\mcitedefaultseppunct}\relax
\EndOfBibitem
\bibitem[Moosm\"{u}ller \emph{et~al.}(1989)Moosm\"{u}ller, She, and
  Huo]{Moosmuller1989}
H.~Moosm\"{u}ller, C.~Y. She and W.~M. Huo, \emph{Phys. Rev. A}, 1989,
  \textbf{40}, 6983--6998\relax
\mciteBstWouldAddEndPuncttrue
\mciteSetBstMidEndSepPunct{\mcitedefaultmidpunct}
{\mcitedefaultendpunct}{\mcitedefaultseppunct}\relax
\EndOfBibitem
\bibitem[Dyer and Bischel(1991)]{Dyer1991}
M.~J. Dyer and W.~K. Bischel, \emph{Phys. Rev. A}, 1991, \textbf{44},
  3138--3143\relax
\mciteBstWouldAddEndPuncttrue
\mciteSetBstMidEndSepPunct{\mcitedefaultmidpunct}
{\mcitedefaultendpunct}{\mcitedefaultseppunct}\relax
\EndOfBibitem
\bibitem[Rosasco \emph{et~al.}(1991)Rosasco, Bowers, Hurst, Looney, Smyth, and
  May]{Rosasco1991}
G.~J. Rosasco, W.~J. Bowers, W.~S. Hurst, J.~P. Looney, K.~C. Smyth and A.~D.
  May, \emph{J. Chem. Phys.}, 1991, \textbf{94}, 7625--7633\relax
\mciteBstWouldAddEndPuncttrue
\mciteSetBstMidEndSepPunct{\mcitedefaultmidpunct}
{\mcitedefaultendpunct}{\mcitedefaultseppunct}\relax
\EndOfBibitem
\bibitem[Rahn and Rosasco(1990)]{Rahn1990}
L.~A. Rahn and G.~J. Rosasco, \emph{Phys. Rev. A}, 1990, \textbf{41},
  3698\relax
\mciteBstWouldAddEndPuncttrue
\mciteSetBstMidEndSepPunct{\mcitedefaultmidpunct}
{\mcitedefaultendpunct}{\mcitedefaultseppunct}\relax
\EndOfBibitem
\bibitem[Rosasco \emph{et~al.}(1989)Rosasco, May, Hurst, Petway, and
  Smyth]{Rosasco1989}
G.~J. Rosasco, A.~D. May, W.~S. Hurst, L.~B. Petway and K.~C. Smyth, \emph{J.
  Chem. Phys.}, 1989, \textbf{90}, 2115--2124\relax
\mciteBstWouldAddEndPuncttrue
\mciteSetBstMidEndSepPunct{\mcitedefaultmidpunct}
{\mcitedefaultendpunct}{\mcitedefaultseppunct}\relax
\EndOfBibitem
\bibitem[Nazemi \emph{et~al.}(1983)Nazemi, Javan, and Pine]{Nazemi1983}
S.~Nazemi, A.~Javan and A.~S. Pine, \emph{J. Chem. Phys.}, 1983, \textbf{78},
  4797--4805\relax
\mciteBstWouldAddEndPuncttrue
\mciteSetBstMidEndSepPunct{\mcitedefaultmidpunct}
{\mcitedefaultendpunct}{\mcitedefaultseppunct}\relax
\EndOfBibitem
\bibitem[Diouf \emph{et~al.}(2019)Diouf, Cozijn, Darqui\'{e}, Salumbides, and
  Ubachs]{Diouf2019}
M.~L. Diouf, F.~M.~J. Cozijn, B.~Darqui\'{e}, E.~J. Salumbides and W.~Ubachs,
  \emph{Opt. Lett.}, 2019, \textbf{44}, 4733\relax
\mciteBstWouldAddEndPuncttrue
\mciteSetBstMidEndSepPunct{\mcitedefaultmidpunct}
{\mcitedefaultendpunct}{\mcitedefaultseppunct}\relax
\EndOfBibitem
\bibitem[Raj \emph{et~al.}(2019)Raj, Witek, and Hamaguchi]{Raj2019}
A.~Raj, H.~A. Witek and H.-O. Hamaguchi, \emph{Mol. Phys.}, 2019, \textbf{0},
  1--13\relax
\mciteBstWouldAddEndPuncttrue
\mciteSetBstMidEndSepPunct{\mcitedefaultmidpunct}
{\mcitedefaultendpunct}{\mcitedefaultseppunct}\relax
\EndOfBibitem
\bibitem[Ramsey and Lewis(1957)]{Ramsey1957}
N.~F. Ramsey and H.~R. Lewis, \emph{Phys. Rev.}, 1957, \textbf{108},
  1246--1250\relax
\mciteBstWouldAddEndPuncttrue
\mciteSetBstMidEndSepPunct{\mcitedefaultmidpunct}
{\mcitedefaultendpunct}{\mcitedefaultseppunct}\relax
\EndOfBibitem
\bibitem[Niu \emph{et~al.}(2014)Niu, Salumbides, Dickenson, Eikema, and
  Ubachs]{Niu2014}
M.~L. Niu, E.~J. Salumbides, G.~D. Dickenson, K.~S.~E. Eikema and W.~Ubachs,
  \emph{J. Mol. Spectrosc.}, 2014, \textbf{300}, 44--54\relax
\mciteBstWouldAddEndPuncttrue
\mciteSetBstMidEndSepPunct{\mcitedefaultmidpunct}
{\mcitedefaultendpunct}{\mcitedefaultseppunct}\relax
\EndOfBibitem
\bibitem[SPE(2019)]{SPECTRE}
H2SPECTRE ver 7.0 Fortran source code, P. Czachorowski, PhD Thesis, University
  of Warsaw, Poland, 2019,
  \url{https://www.qcg.home.amu.edu.pl/qcg/public\_html/H2Spectre.html}\relax
\mciteBstWouldAddEndPuncttrue
\mciteSetBstMidEndSepPunct{\mcitedefaultmidpunct}
{\mcitedefaultendpunct}{\mcitedefaultseppunct}\relax
\EndOfBibitem
\bibitem[Lucht and Farrow(1989)]{Lucht1989}
R.~P. Lucht and R.~L. Farrow, \emph{J. Opt. Soc. Am. B}, 1989, \textbf{6},
  2313--2325\relax
\mciteBstWouldAddEndPuncttrue
\mciteSetBstMidEndSepPunct{\mcitedefaultmidpunct}
{\mcitedefaultendpunct}{\mcitedefaultseppunct}\relax
\EndOfBibitem
\bibitem[Owyoung and Esherick(1980)]{Owyoung1980}
A.~Owyoung and P.~Esherick, \emph{Opt. Lett.}, 1980, \textbf{5}, 421--423\relax
\mciteBstWouldAddEndPuncttrue
\mciteSetBstMidEndSepPunct{\mcitedefaultmidpunct}
{\mcitedefaultendpunct}{\mcitedefaultseppunct}\relax
\EndOfBibitem
\bibitem[Bombach \emph{et~al.}(1990)Bombach, Hemmerling, and
  Hubschmid]{Bombach1990}
R.~Bombach, B.~Hemmerling and W.~Hubschmid, \emph{Chem. Phys.}, 1990,
  \textbf{144}, 265--271\relax
\mciteBstWouldAddEndPuncttrue
\mciteSetBstMidEndSepPunct{\mcitedefaultmidpunct}
{\mcitedefaultendpunct}{\mcitedefaultseppunct}\relax
\EndOfBibitem
\bibitem[Trivikram \emph{et~al.}(2016)Trivikram, Niu, Wcis{\l}o, Ubachs, and
  Salumbides]{Trivikram2016}
T.~M. Trivikram, M.~L. Niu, P.~Wcis{\l}o, W.~Ubachs and E.~J. Salumbides,
  \emph{Appl. Phys. B}, 2016, \textbf{122}, 294\relax
\mciteBstWouldAddEndPuncttrue
\mciteSetBstMidEndSepPunct{\mcitedefaultmidpunct}
{\mcitedefaultendpunct}{\mcitedefaultseppunct}\relax
\EndOfBibitem
\bibitem[Li \emph{et~al.}(1985)Li, Yang, and Johnson]{Li1985}
L.~Li, B.-X. Yang and P.~M. Johnson, \emph{J. Opt. Soc. Am. B}, 1985,
  \textbf{2}, 748--752\relax
\mciteBstWouldAddEndPuncttrue
\mciteSetBstMidEndSepPunct{\mcitedefaultmidpunct}
{\mcitedefaultendpunct}{\mcitedefaultseppunct}\relax
\EndOfBibitem
\bibitem[Farrow and Rahn(1982)]{Farrow1982}
R.~L. Farrow and L.~A. Rahn, \emph{Phys. Rev. Lett.}, 1982, \textbf{48},
  395--398\relax
\mciteBstWouldAddEndPuncttrue
\mciteSetBstMidEndSepPunct{\mcitedefaultmidpunct}
{\mcitedefaultendpunct}{\mcitedefaultseppunct}\relax
\EndOfBibitem
\bibitem[Lundeen \emph{et~al.}(1983)Lundeen, Hou, and Nibler]{Lundeen1983}
T.~Lundeen, S.~Y. Hou and J.~W. Nibler, \emph{J. Chem. Phys.}, 1983,
  \textbf{79}, 6301--6305\relax
\mciteBstWouldAddEndPuncttrue
\mciteSetBstMidEndSepPunct{\mcitedefaultmidpunct}
{\mcitedefaultendpunct}{\mcitedefaultseppunct}\relax
\EndOfBibitem
\bibitem[Druet \emph{et~al.}(1979)Druet, Taran, and Bord\'{e}]{Druet1979}
S.~Druet, J.-P. Taran and C.~J. Bord\'{e}, \emph{J. Phys. France}, 1979,
  \textbf{40}, 819--840\relax
\mciteBstWouldAddEndPuncttrue
\mciteSetBstMidEndSepPunct{\mcitedefaultmidpunct}
{\mcitedefaultendpunct}{\mcitedefaultseppunct}\relax
\EndOfBibitem
\bibitem[Rosasco and Hurst(1985)]{Rosasco1985}
G.~J. Rosasco and W.~S. Hurst, \emph{Phys. Rev. A}, 1985, \textbf{32},
  281--299\relax
\mciteBstWouldAddEndPuncttrue
\mciteSetBstMidEndSepPunct{\mcitedefaultmidpunct}
{\mcitedefaultendpunct}{\mcitedefaultseppunct}\relax
\EndOfBibitem
\bibitem[Raj \emph{et~al.}(2018)Raj, Hamaguchi, and Witek]{Raj2018}
A.~Raj, H.-O. Hamaguchi and H.~A. Witek, \emph{J. Chem. Phys.}, 2018,
  \textbf{148}, 104308\relax
\mciteBstWouldAddEndPuncttrue
\mciteSetBstMidEndSepPunct{\mcitedefaultmidpunct}
{\mcitedefaultendpunct}{\mcitedefaultseppunct}\relax
\EndOfBibitem
\bibitem[Hahn and Lee(1995)]{Hahn1995}
J.~W. Hahn and E.~S. Lee, \emph{J. Opt. Soc. Am. B}, 1995, \textbf{12},
  1021\relax
\mciteBstWouldAddEndPuncttrue
\mciteSetBstMidEndSepPunct{\mcitedefaultmidpunct}
{\mcitedefaultendpunct}{\mcitedefaultseppunct}\relax
\EndOfBibitem
\end{mcitethebibliography}
\bibliographystyle{rsc} 

\end{document}